\documentclass[10pt]{article}
\usepackage{graphicx}
\usepackage{cite}

\begin{document} 

\title{Universal low-temperature behavior 
       of frustrated quantum antiferromagnets
       in the vicinity of the saturation field}

\author{Oleg Derzhko$^{1,2}$ and Johannes Richter$^2$\\
\small {$^1$Institute for Condensed Matter Physics,
            National Academy of Sciences of Ukraine,}\\
\small {1 Svientsitskii Street, L'viv-11, 79011, Ukraine}\\
\small {$^2$Institut f\"{u}r Theoretische Physik,
            Universit\"{a}t Magdeburg,}\\
\small{P.O. Box 4120, D-39016 Magdeburg, Germany}}

\date{\today}
       
\maketitle

\renewcommand\baselinestretch {1.2}
\large\normalsize

\begin{abstract}
We study the low-temperature thermodynamic properties 
of a number of frustrated quantum antiferromagnets 
which support localized magnon states
in the vicinity of the saturation field.
For this purpose we use 
1) a mapping of the low-energy degrees of freedom of spin systems 
onto the hard-core object lattice gases
and 
2) an exact diagonalization of finite spin systems 
of up to $N=30$ sites.
The considered spin systems exhibit universal behavior
which is determined by a specific hard-core object lattice gas
representing the independent localized magnon states.
We test the lattice gas description 
by comparing its predictions 
with the numerical results 
for low-lying energy states of finite spin systems. 
For all frustrated spin systems considered we find a strong variation 
of the low-temperature specific heat passing the saturation field and a
maximum in the isothermal entropy at saturation field resulting in an
enhanced magnetocaloric effect.
\end{abstract}


75.45.+j Macroscopic quantum phenomena in magnetic systems

75.30.Sg Magnetocaloric effect, magnetic cooling

\renewcommand\baselinestretch {1.31}
\large\normalsize

\section{Introductory remarks}
\label{1}

Quantum spin antiferromagnets on geometrically frustrated lattices 
have attracted much attention during last years
\cite{01,02,03}.
A new and rapidly developing direction 
in this field of quantum magnetism
is a study of the low-temperature properties of such systems 
in the presence of an external magnetic field. 
The Zeeman interaction of spins with a magnetic field 
competes with the frustrating antiferromagnetic interspin interactions
that may lead to new phenomena.
Recently it has been found 
that a wide class of geometrically frustrated quantum spin antiferromagnets
(including the kagom\'{e} and pyrochlore antiferromagnets)
has quite simple ground states 
in the vicinity of the saturation magnetic field \cite{04,05,06}
(for a review, see Refs. \cite{07,08}).
These ground states consist of independent (i.e. isolated) localized magnons 
in a ferromagnetic environment.
The localized magnon states were used to predict  
a ground-state magnetization jump at the saturation field
\cite{04,05,06},
a magnetic field induced spin-Peierls instability
\cite{09,10},
and
a residual ground-state entropy at the saturation field
\cite{03,11,12,13,08}.
Moreover, 
in Refs. \cite{12,13,08} 
the concept of localized magnons was used for a detailed analysis 
of the low-temperature magnetothermodynamics 
in the vicinity of the saturation field 
for two representative systems, 
the sawtooth chain
(or $\Delta$-chain)
and 
the kagom\'{e} lattice.
In particular, 
the authors of these papers mapped the low-energy degrees of freedom 
of the sawtooth chain 
(the kagom\'{e} lattice) 
to the hard-dimer gas on a one-dimensional lattice 
(the hard-hexagon gas on a triangular lattice) 
and used the results for the classical lattice gases 
to discuss the properties of the spin systems.
They also provided exact diagonalization data for finite sawtooth chains 
(up to $N=20$ sites)
to illustrate the efficiency of the hard-dimer description 
of the spin chain 
at low temperatures near the saturation field.

In the present paper 
we extend the previous studies 
on the low-temperature strong-field magnetothermodynamics 
examining various other frustrated quantum spin antiferromagnets
supporting localized magnon states.
We emphasize 
that all such spin systems exhibit a universal behavior. 
It is determined by a specific hard-core object lattice gas 
which mimics the independent localized magnon states.
Thus,
the one-dimensional hard-dimer behavior 
is also inherent in the two-leg ladder or kagom\'{e}-like chains, 
whereas
the hard-hexagon behavior is also inherent in the star lattice
(see Table \ref{tab1} below).
Moreover,
we examine the thermodynamic properties of the spin models 
which are described by a gas of monomers.
We also provide exact diagonalization data for finite spin systems 
of up to $N=30$ sites 
and discuss to what extent a hard-core object description 
can reproduce the properties of the spin systems.
We compare analytical and numerical results 
for different one-dimensional lattices 
and for the two-dimensional square-kagom\'{e} lattice.

The remainder of the paper is organized as follows.
In Sec.~\ref{2} 
we introduce the spin models to be discussed.
In Sec.~\ref{3} 
we discuss a hard-core object description 
of the independent localized magnon states. 
We also consider how the lattice gas description can be extended 
for a wider temperature/field region.
In Sec.~\ref{4} 
we present the exact diagonalization data for finite systems 
and compare them with the analytical predictions 
which follow from the hard-core object picture.
We discuss briefly 
the effects of the localized magnons
on the temperature dependence of the specific heat 
and the magnetocaloric effect near the saturation field.
Finally,
in Sec.~\ref{5}
we summarize our findings.

\section{Geometrically frustrated spin models and localized magnons}
\label{2}

\begin{figure}
\begin{center}
\includegraphics[clip=on,width=50mm,angle=0]{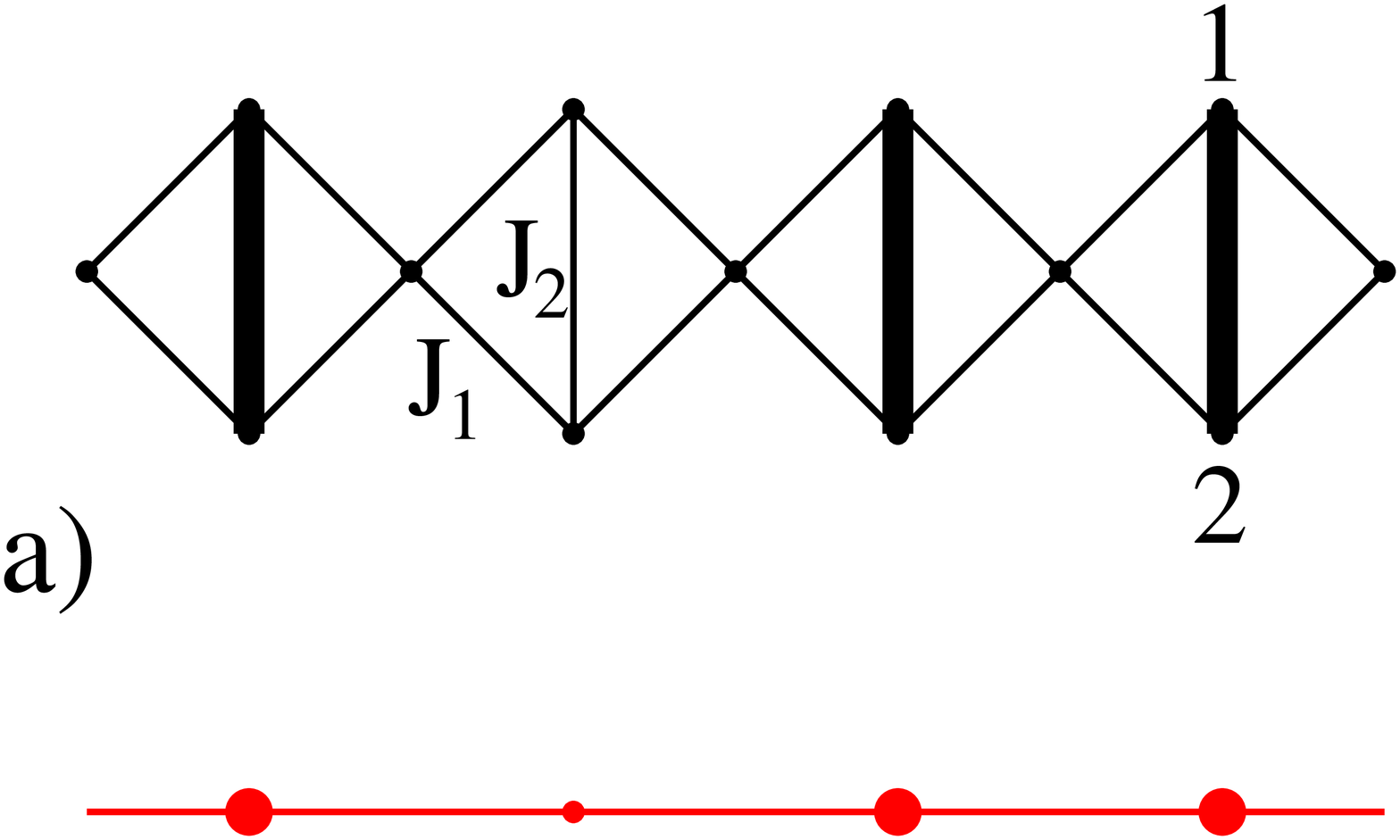}
\hspace{5mm}
\includegraphics[clip=on,width=90mm,angle=0]{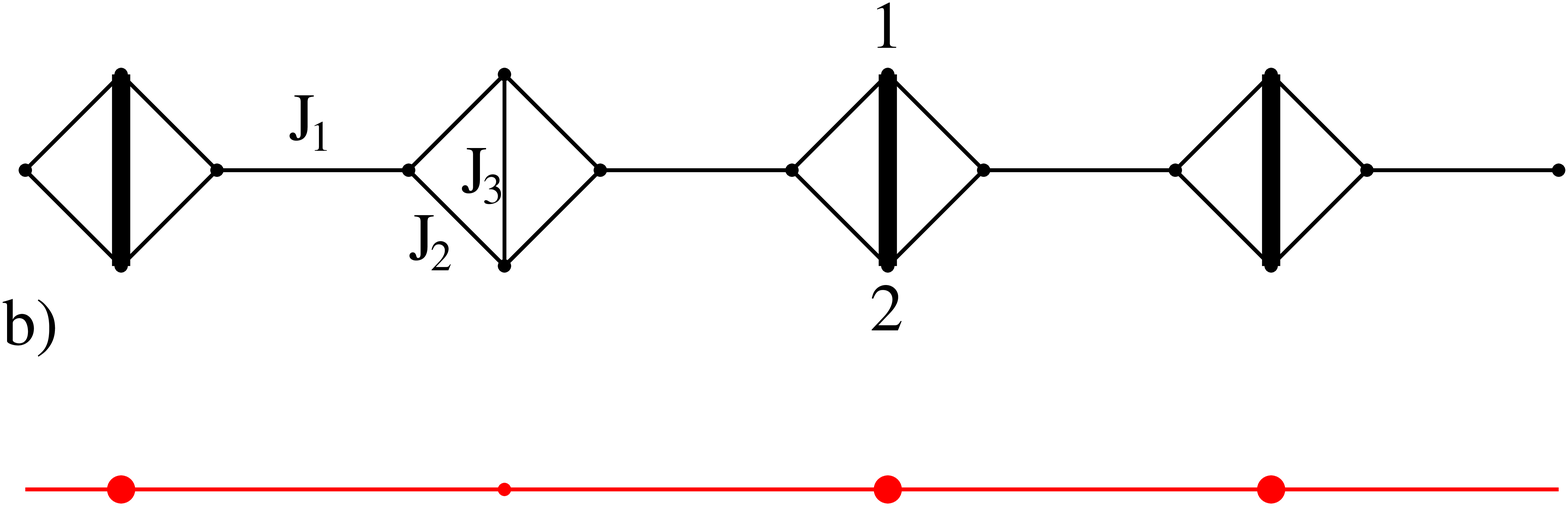}
\\
\vspace{20mm}
\includegraphics[clip=on,width=75mm,angle=0]{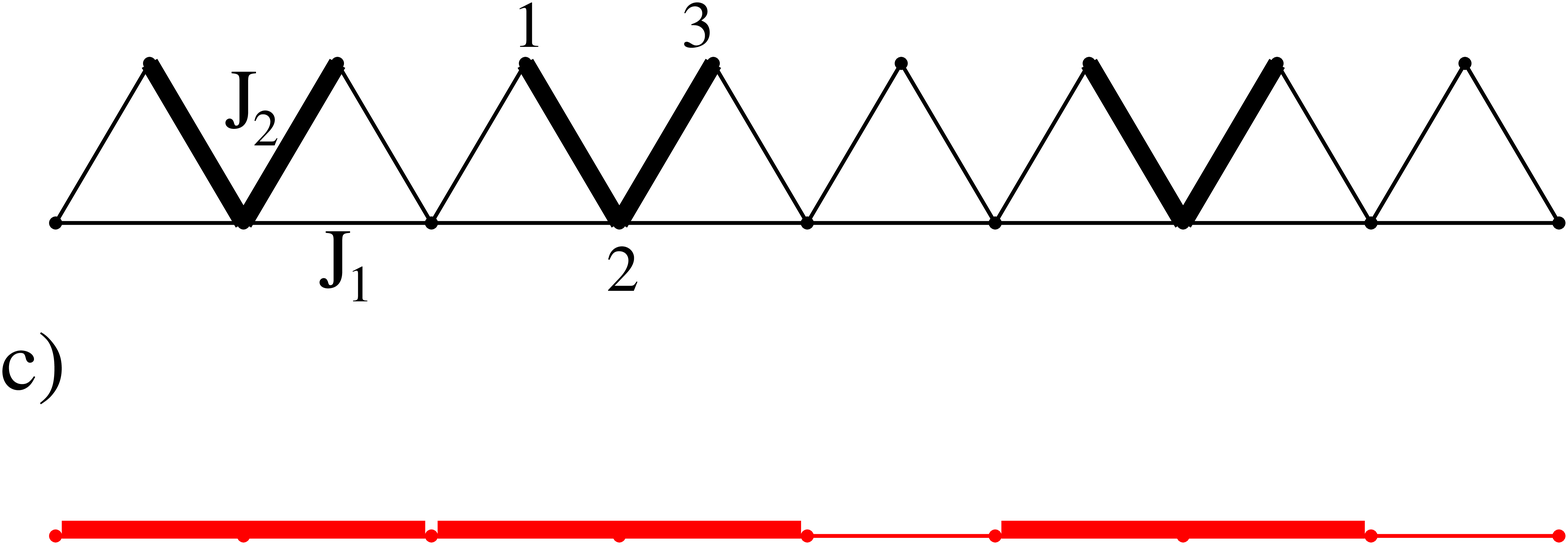}
\hspace{5mm}
\includegraphics[clip=on,width=70mm,angle=0]{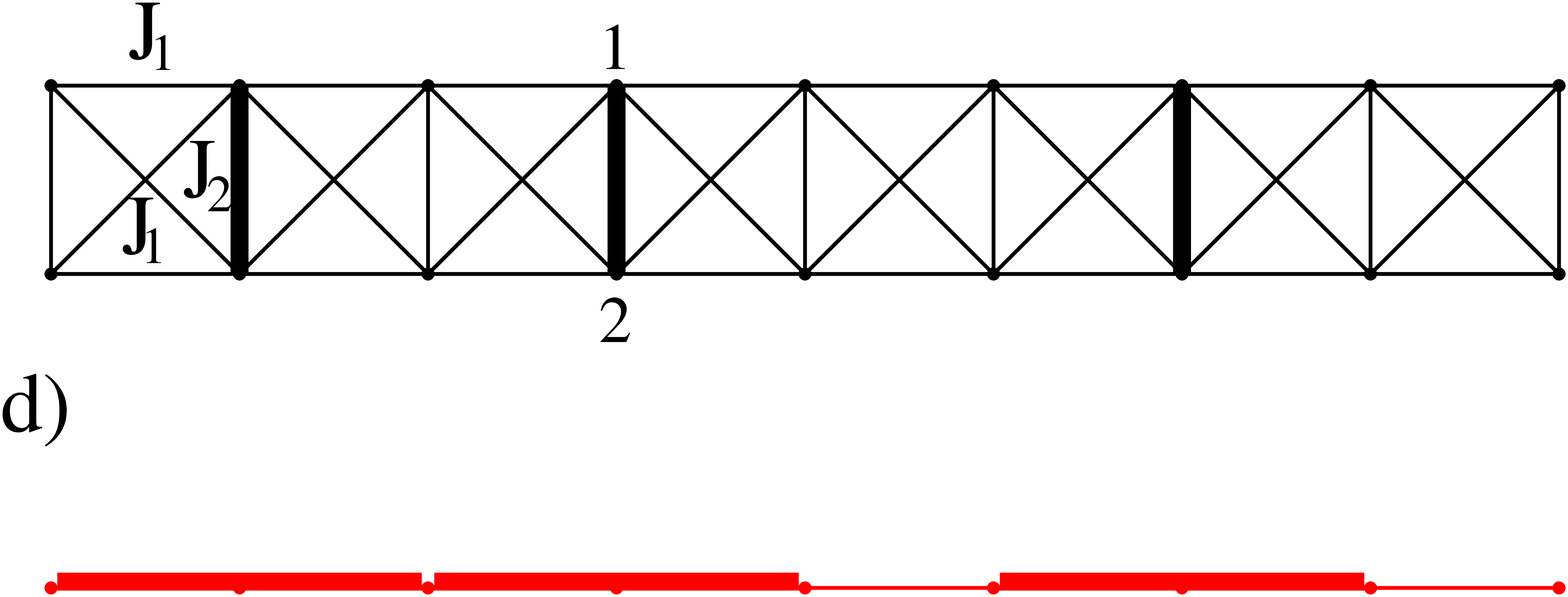}
\\
\vspace{20mm}
\includegraphics[clip=on,width=65mm,angle=0]{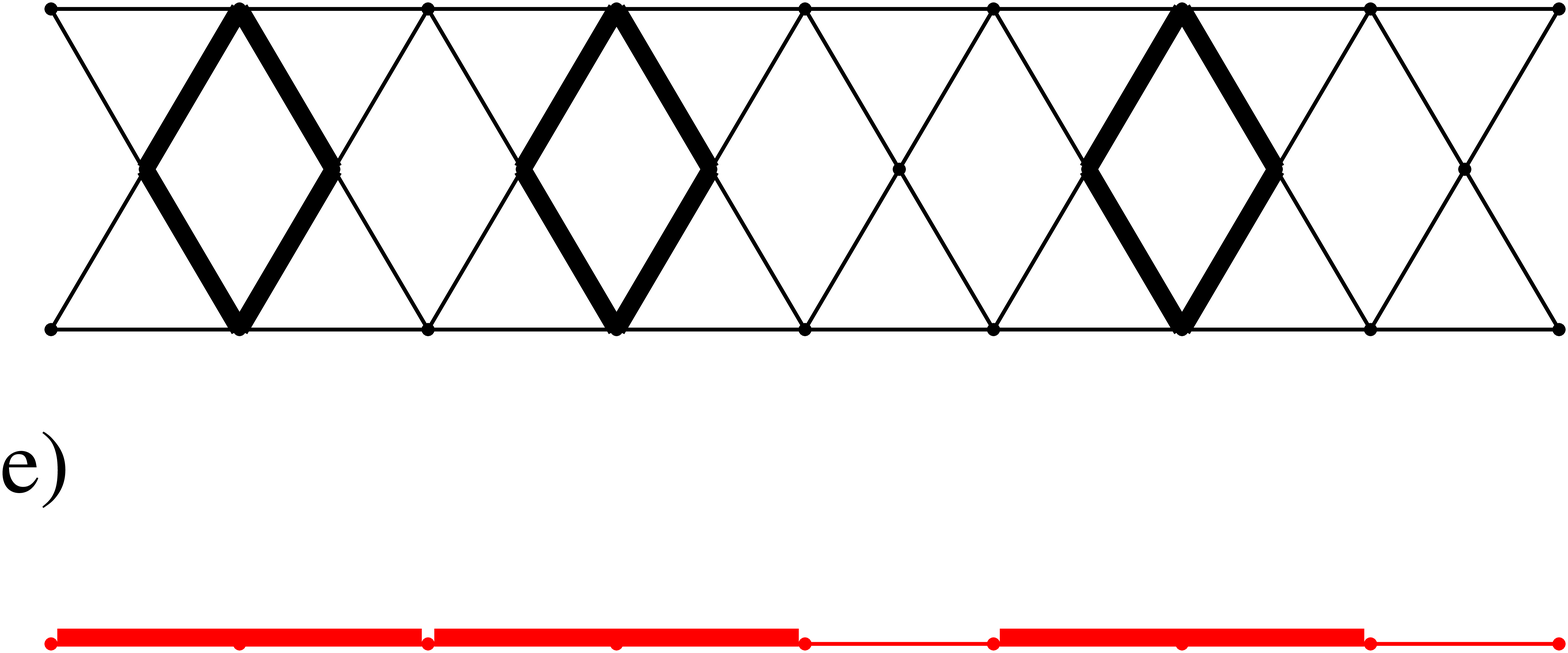}
\hspace{5mm}
\includegraphics[clip=on,width=70mm,angle=0]{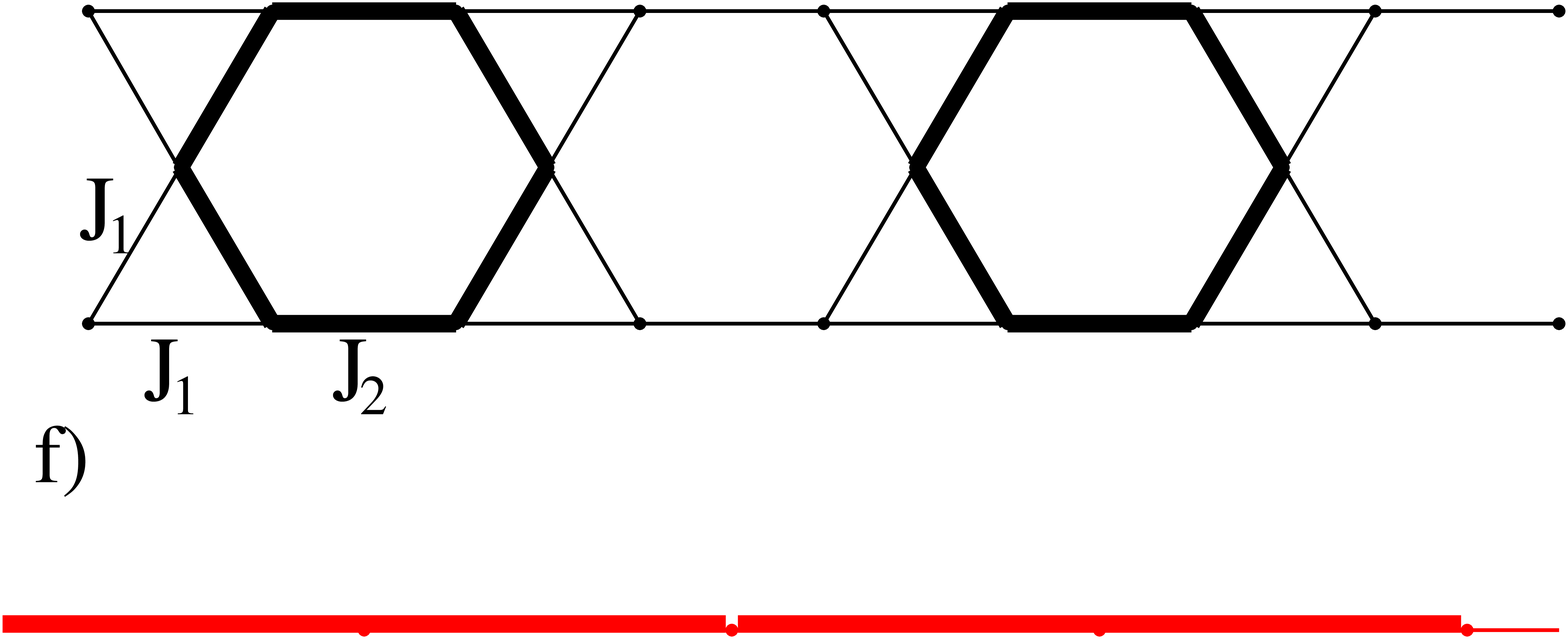}
\\
\vspace{20mm}
\end{center}
\caption[]
{(Color online)
Several one-dimensional frustrated quantum spin lattices 
supporting localized magnon states:
the diamond chain \cite{14} (a),
the dimer-plaquette chain \cite{15} (b),
the sawtooth chain \cite{16} (c),
the two-leg ladder 
(the spins are sitting only on the squares, 
not on the intersections of the diagonals)
\cite{17} (d),
and
two kagom\'{e}-like chains \cite{18,19} (e and f).
The trapping cells occupied by localized magnons are shown by fat lines.
Below each lattice we show the corresponding auxiliary lattice 
filled by hard-core objects
(monomers (a and b) and dimers (c, d, e and f)).
\label{fig1}}
\end{figure}

\begin{figure}
\begin{center}
\includegraphics[clip=on,width=5cm,angle=0]{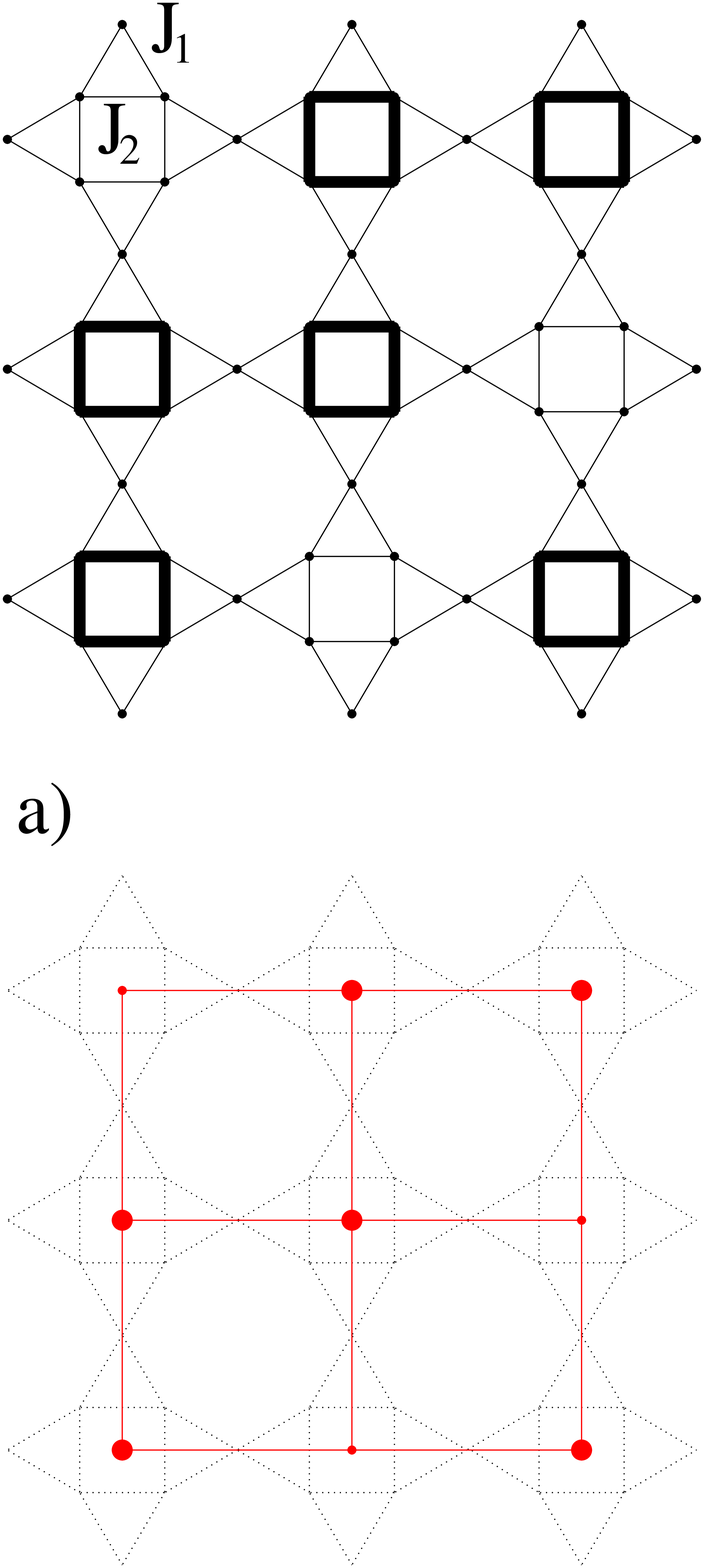}
\hspace{25mm}
\includegraphics[clip=on,width=5cm,angle=0]{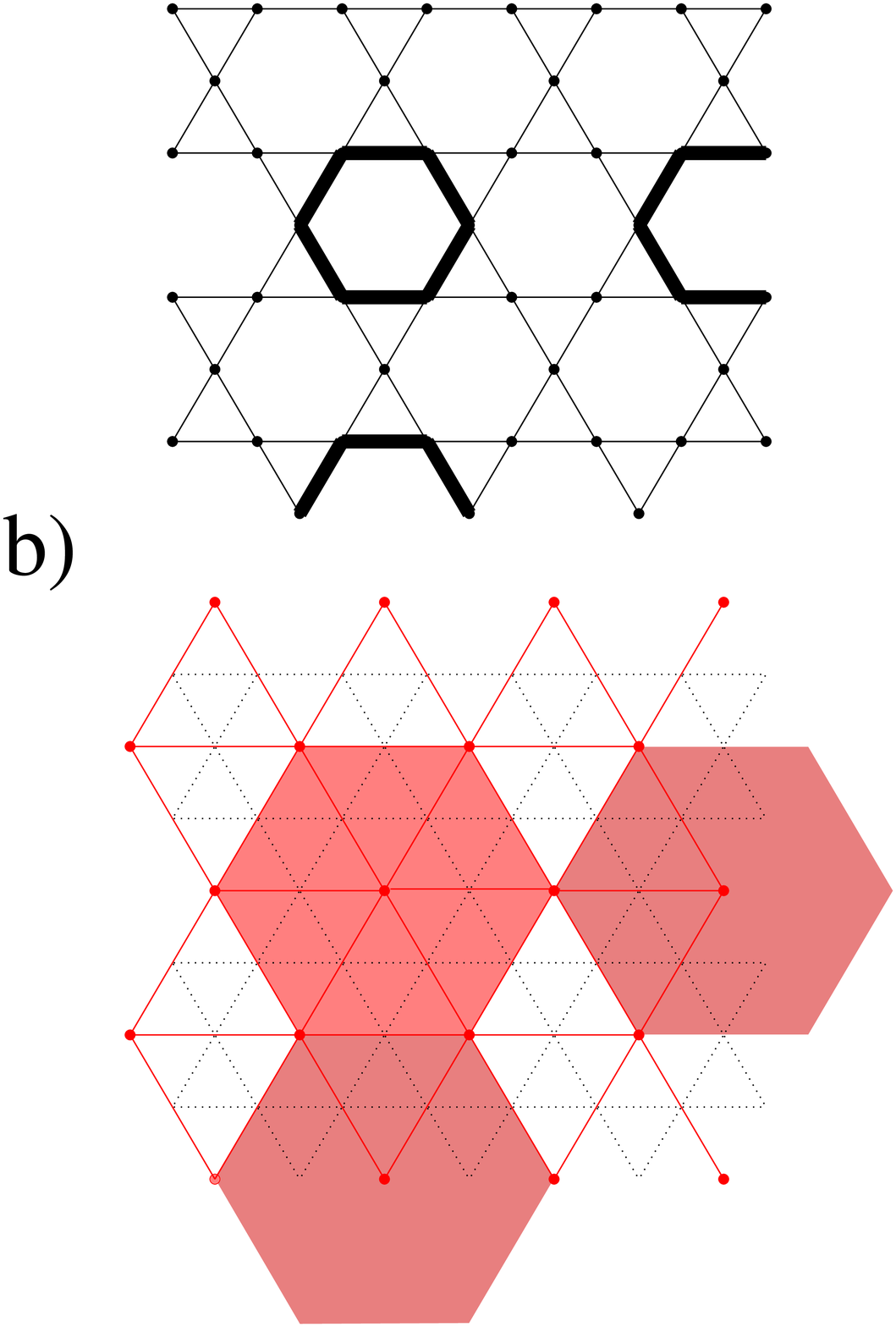}
\\
\vspace{20mm}
\includegraphics[clip=on,width=5cm,angle=0]{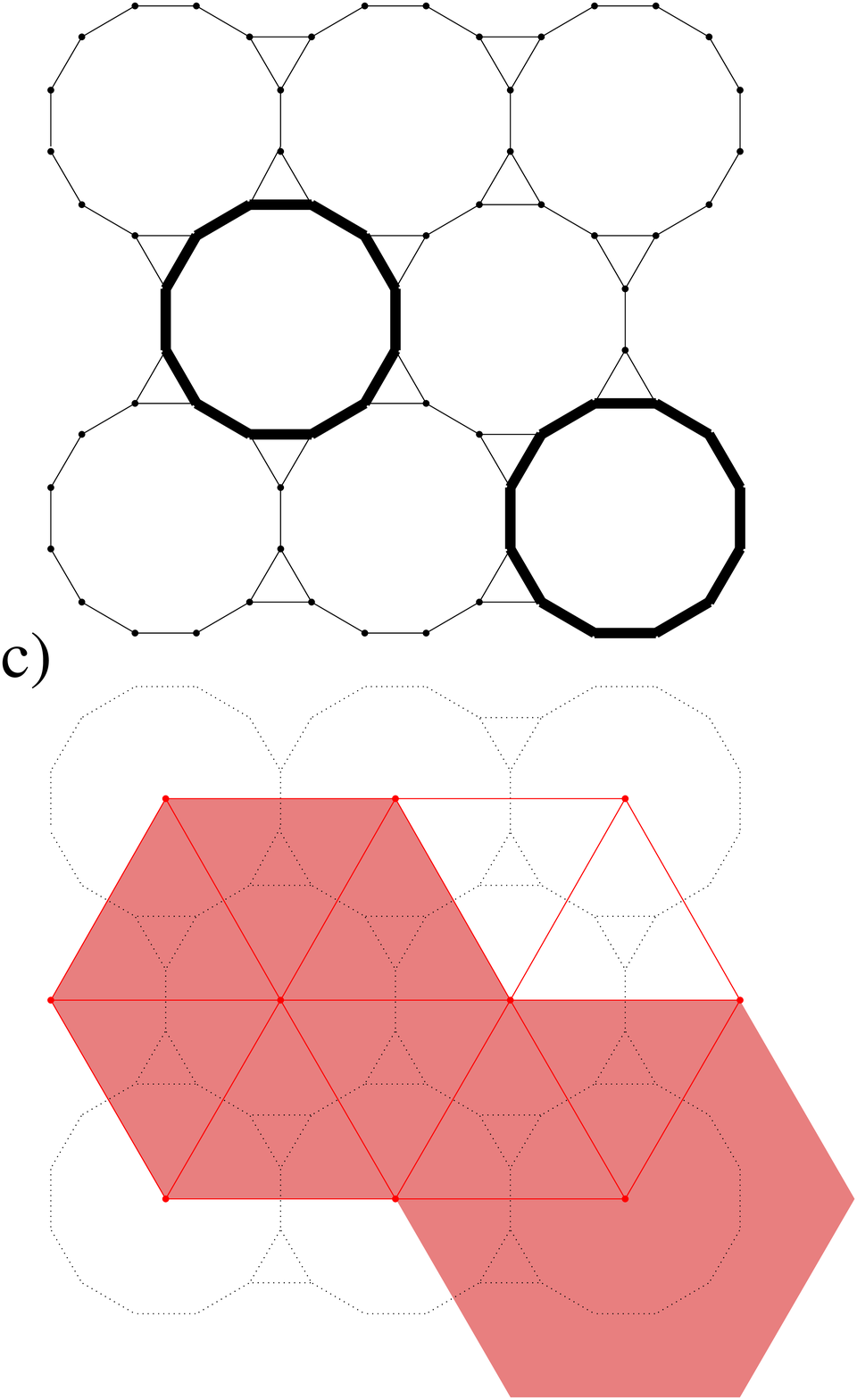}
\hspace{25mm}
\includegraphics[clip=on,width=5cm,angle=0]{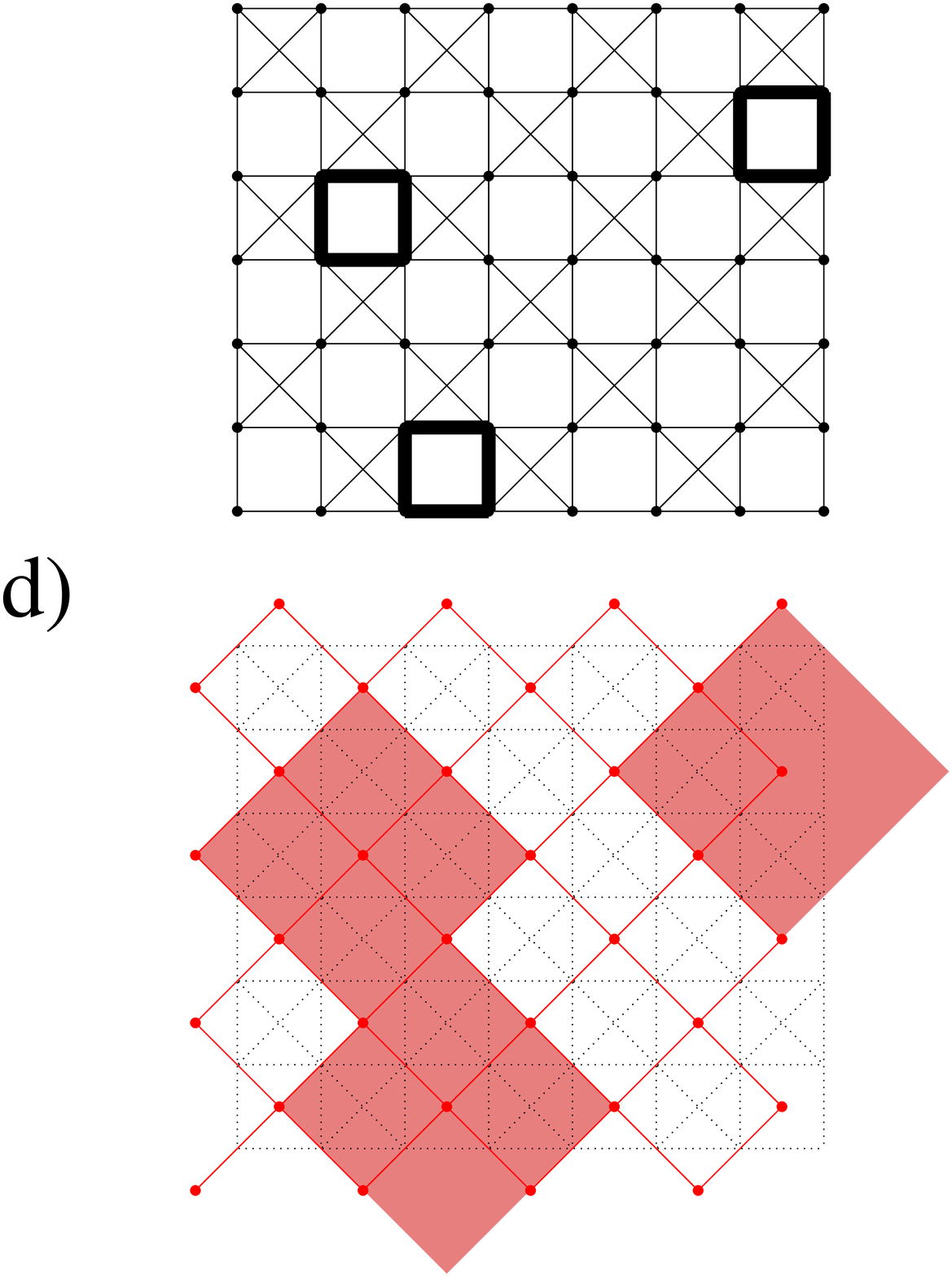}
\end{center}
\caption[]
{(Color online) 
Several two-dimensional frustrated quantum spin lattices 
supporting localized magnon states:
the square-kagom\'{e} lattice \cite{20} (a),
the kagom\'{e} lattice \cite{03,05,21} (b),
the star lattice \cite{03,22} (c),
and
the checkerboard lattice 
(the spins are sitting only on the squares, 
not on the intersections of the diagonals)
\cite{23} (d).
The trapping cells occupied by localized magnons are shown by fat lines.
Below each lattice we show the corresponding auxiliary lattice
(square (a and d) and triangular (b and c) lattices) 
filled by hard-core objects
(monomers (a), hexagons (b and c) and squares (d)).
\label{fig2}}
\end{figure}

In this paper
we consider several frustrated quantum spin lattices
discussed so far in the literature by various authors,
namely,
the diamond chain \cite{14}, 
the dimer-plaquette chain \cite{15}, 
the sawtooth chain \cite{16},
the two-leg ladder \cite{17},
two kagom\'{e}-like chains \cite{18,19}
(Fig. \ref{fig1}),
the square-kagom\'{e} lattice \cite{20},
the kagom\'{e} lattice \cite{03,05,21},
the star lattice \cite{03,22},
the checkerboard lattice \cite{23}
(Fig. \ref{fig2}).
(Note that further models hosting localized magnons can be constructed.)
The Hamiltonian of the system consisting of $N$ antiferromagnetically interacting spins in a magnetic 
field $h$ reads
\begin{eqnarray}
H
=\sum_{(ij)}J_{ij}{\bf{s}}_i\cdot{\bf{s}}_j-hS^z
=\sum_{(ij)}J_{ij}\left(\frac{1}{2}\left(s^+_i s^-_j+s^-_i s^+_j\right)
+s^z_i s^z_j\right)
-hS^z,
\;\;\;
S^z=\sum_i s_i^z  
\label{01}
\end{eqnarray}
with ${\bf s}_i^2= s(s+1)$ and $ J_{ij}>0$.
From Refs. \cite{04,05,06,07,03} we know 
that for certain relations between $J_{ij}$ 
(see Table \ref{tab1})
all spin lattices shown in Figs.~\ref{fig1}  and \ref{fig2}
support localized magnon states
which become relevant at  strong magnetic fields.
In what follows we fix the energy scale by setting either $J=1$ (for 
systems with only one exchange integral) or $J_1=1$ (for 
systems with more then one exchange integral). 
\begin{table}
\begin{center}
\caption
{Specific and generic properties of the spin systems 
shown in Figs. \ref{fig1} and \ref{fig2}.
$\epsilon_1$ is the difference 
between the energy of the fully polarized state 
and the energy of the one-magnon state 
in zero field and determines the saturation field $h_1=\epsilon_1$,
$n_{\max}$ is the maximum number of independent localized magnons 
for the closest possible packing,
${\cal{N}}$ is the number of sites in the auxiliary lattice 
with hard-core objects,
$N$ is the number of sites in the spin lattice. 
\label{tab1}}
\vspace{5mm}
\begin{tabular}{|c||c|c|c|c|c|} \hline
lattice & relation between $J_{ij}$ & $\epsilon_1$ & $n_{\max}/N$ & ${\cal{N}}/N$ 
& universality class \\ \hline \hline 
Fig. \ref{fig1}a
        & $J_2\ge 2J_1$
                                    & $2s(J_1+J_2)$
                                                   & 1/3
                                                                  & 1/3  
& monomers           \\ \hline
Fig. \ref{fig1}b
        & $J_3\ge J_3^c(J_1,J_2)$;
                                    & $2s(J_1+J_3)$ 
                                                   & 1/4
                                                                  & 1/4                                                                                             
& monomers           \\ 
         & if $J_1=J_2$ then $J_3\ge ((1+\sqrt{5})/2) J_1$
                                    &
                                                   &
                                                                  &
&                    \\ \hline
Fig. \ref{fig1}c
        & $J_2=2J_1$
                                    & $8sJ_1$ 
                                                   & 1/4
                                                                  & 1/2           
& dimers             \\ \hline
Fig. \ref{fig1}d
        & $J_2\ge 2J_1$
                                    & $2s(2J_1+J_2)$ 
                                                   & 1/4
                                                                  & 1/2           
& dimers             \\ \hline
Fig. \ref{fig1}e
        & all bonds have the same strength $J$
                                    & $6sJ$
                                                   & 1/6
                                                                  & 1/3           
& dimers             \\ \hline
Fig. \ref{fig1}f
        & $J_2=(3/2)J_1$
                                    & $6sJ_1$
                                                   & 1/10
                                                                  & 1/5           
& dimers             \\ \hline
Fig. \ref{fig2}a
        & $J_2\ge J_1$
                                    & $2s(J_1+2J_2)$ 
                                                   & 1/6
                                                                  & 1/6           
& monomers           \\ \hline
Fig. \ref{fig2}b
        & all bonds have the same strength $J$
                                    & $6sJ$ 
                                                   & 1/9
                                                                  & 1/3           
& hexagons           \\ \hline
Fig. \ref{fig2}c
        & all bonds have the same strength $J$
                                    & $5sJ$ 
                                                   & 1/18
                                                                  & 1/6           
& hexagons           \\ \hline
Fig. \ref{fig2}d
        & all bonds have the same strength $J$
                                    & $8sJ$ 
                                                   & 1/8
                                                                  & 1/2           
& squares            \\ \hline
\end{tabular}
\end{center}
\end{table}
For the diamond and dimer-plaquette chains and for the two-leg ladder 
the magnons may be localized on the vertical bonds 
(see panels a, b and d in Fig. \ref{fig1}),
for the sawtooth chain the localized magnons may be trapped 
in a valley between two neighboring triangles
(panel c in Fig. \ref{fig1}),
whereas for other lattices 
the localized magnons may occupy the even polygons 
shown by fat lines in Figs. \ref{fig1}, \ref{fig2}.
The explicit expressions for the localized one-magnon state 
having the smallest possible region of localization  
read
\begin{eqnarray}
\vert 1{\rm{lm}}\rangle 
\propto
\left(\vert{s-1}_1,{s}_2\rangle-\vert{s}_1,{s-1}_2\rangle\right)\vert s,\ldots,s\rangle
\nonumber
\end{eqnarray}
(panels a, b and d in Fig. \ref{fig1}),
\begin{eqnarray}
\vert 1{\rm{lm}}\rangle 
\propto
\left(\vert{s-1}_1,{s}_2,{s}_3\rangle
-2\vert{s}_1,{s-1}_2,{s}_3\rangle
+\vert{s}_1,{s}_2,{s-1}_3\rangle\right)
\vert s,\ldots,s\rangle
\nonumber
\end{eqnarray}
(panel c in Fig. \ref{fig1}),
\begin{eqnarray}
\vert 1{\rm{lm}}\rangle 
\propto
\sum_m(-1)^{m}s_m^{-}\vert s,\ldots,s\rangle
\nonumber
\end{eqnarray}
(the sum over $m$ runs, say, counterclockwise over the sites $m$ of a trapping polygon 
in panels e and f in Fig. \ref{fig1} and in all panels in Fig. \ref{fig2}).
Here $s_i$ or ${s-1}_i$ denote the values of $s_i^z$
and $s$ is the maximal possible value of $s_i^z$.  
We start with the zero-field case.
The energy of the localized one-magnon state 
can be written as
$E_{\rm{FM}} -\epsilon_1$,
where $E_{\rm{FM}}$ 
is the energy of the fully (ferromagnetically) polarized spin system
and the values of $\epsilon_1$ for the considered systems 
are given in Table \ref{tab1}.
We can fill the spin lattice 
by $n=1,\ldots,n_{\max}$, $n_{\max}\propto N$ (see Table \ref{tab1}) 
independent localized magnons.
Each localized magnon decreases the magnetization $S^z$ by 1
and the state with $n$ localized magnons has $S^{z}=Ns-n$.
The independent localized magnon states are the lowest-energy states 
in the corresponding sectors of 
$S^z=Ns-1,\ldots,Ns-n_{\max}$ \cite{04,24}. 
Although the localized magnon states exist also
for 
anisotropic $XXZ$ Heisenberg exchange interaction in (\ref{01}) 
and 
arbitrary spin length $s$,
in what follows we restrict ourselves 
to the isotropic (i.e. $XXX$) Heisenberg exchange interaction
and (in numerical computations)
to the extreme quantum case $s=1/2$
without loss of generality for discussion 
of the universal low-temperature behavior of the spin systems 
near the saturation field. 
However, 
we have to bear in mind 
that the effects of the localized magnon states 
are true quantum effects 
which become less and less pronounced as $s\to\infty$ \cite{05,11}.

Consider now the localized magnon states 
in the presence of an external field $h$.
A localized one-magnon state has the energy
$E_1(h)
=
E_{\rm{FM}}-hsN -\epsilon_1+h$,
whereas a state 
with $n$ independent localized magnons
has the energy
\begin{eqnarray}
E_n(h)
=
E_{\rm{FM}}-hsN -n\left(\epsilon_1-h\right).
\label{02}
\end{eqnarray}
Evidently,
$n$ independent localized magnons
can be placed on a lattice with $N$ sites in many ways.
We denote by $g_N(n)\ge 1$ 
the number of ways in which 
$n$ independent localized magnons, 
each occupying the smallest possible area, 
can be put 
on a lattice with $N$ sites.
The energy of any of these $g_N(n)$ states is the same 
and is given by Eq. (\ref{02}),
i.e. $g_N(n)$  
is the degeneracy 
of the independent localized $n$-magnon states.  
At the saturation field $h_1=\epsilon_1$ 
the energy $E_n(h_1)$ (\ref{02}) does not depend on $n$
that further increases the degeneracy 
of the independent localized magnon states.

Let us briefly recall 
the consequences of the localized magnons 
\cite{03,04,05,06,07,08,09,10,11,12,13}.
Due to the localized magnons
different spin systems have identical/universal behavior 
in the ground state in strong magnetic fields. 
Since the lowest-energy states 
with different numbers of independent localized magnons 
$n=1,\ldots,n_{\max}$ have the same energy at the saturation field $h_1$,
$E_{\rm{FM}}-h_1sN$,
the ground-state magnetization exhibits a jump at $h=h_1$
between the values $Ns$ and $Ns-n_{\max}$. This jump is accompanied by a preceding plateau. 
Next, the ground-state energy at the saturation field  
exhibits a huge degeneracy  
which grows exponentially with the system size
(at least as $2^{n_{\max}}$)
and hence the ground-state entropy per site remains finite 
at $h=h_1$.
Finally, 
a lattice distortion 
which preserves the symmetry of the cell which hosts a localized magnon 
may lower the total ground-state energy
(which consists of the magnetic and elastic parts).
This deformation, 
which obviously cannot exist for $h>h_1$, leads to a field-tuned ground-state 
structural instability  in the vicinity of the saturation field.

One may expect 
that the contribution of localized magnons is dominant also 
at low temperatures and strong magnetic fields around the saturation field. 
The analysis for the sawtooth chain and the kagom\'{e} lattice 
reported in Refs. \cite{12,13,08} supports this expectation.
We now move on to discuss the contribution 
of the independent localized magnons  
to the low-temperature properties of the considered spin systems 
in the vicinity of the saturation field.

\section{Independent localized magnons and classical lattice gases}
\label{3}

In this Section,
we examine the contribution of the independent localized magnons 
to the thermodynamic quantities of the spin systems.
Moreover,
we use for this purpose 
a hard-core object representation 
of the independent localized magnons of the smallest area. 
Such a representation allows us to utilize 
the broad knowledge on classical lattice gases.
In the context of this approach 
a number of questions naturally arise 
which we have to consider.
First, 
one has to show 
that the independent localized magnon states of smallest area 
are linearly independent 
in each sectors of $S^z=Ns-1,\ldots,Ns-n_{\max}$.
Only in this case they all contribute 
to the partition function of the spin system.
Although there is no proof 
of the linear independence of the independent localized magnons 
in the general case,
this question has been discussed in Refs. \cite{13,08}
and a rigorous proof for all spin systems 
shown in Figs. \ref{fig1}, \ref{fig2}
is given in Ref. \cite{25}.
Thus, it was shown 
that for the diamond chain, the dimer-plaquette chain, the two-leg ladder, 
and the square-kagom\'{e} lattice 
(the orthogonal type of frustrated spin systems
in the nomenclature of Ref. \cite{25})
as well as for the sawtooth chain and two kagom\'{e}-like chains
(the isolated type of frustrated spin systems) 
the independent localized magnon states 
are linearly independent in every sector of $S^z=Ns-1,\ldots,Ns-n_{\max}$.
Moreover, for a system of the orthogonal type 
they form an orthogonal basis.
The kagom\'{e} lattice, the star lattice and the checkerboard lattice 
belong to the frustrated spin systems 
of codimension one type.
They possess exactly one nontrivial linear relation 
between their localized one-magnon states,
but the set of $n$ independent localized magnon states 
is linearly independent for $n=2,\ldots,n_{\max}$ \cite{25}.
Thus, in the thermodynamic limit
the linear independence of the localized magnons 
is fulfilled for all considered spin systems.

Next,
we have to check that there are no other low-energy states 
(except the independent localized magnons)
in every sector of $S^z=Ns-1,\ldots,Ns-n_{\max}$,
or at least to show, 
if they do exist, 
that their contribution is vanishingly small in the thermodynamic limit.
In Refs. \cite{13,08} 
it was shown that for the kagom\'{e} lattice 
there are indeed a few classes of states 
in the sectors $S^z=Ns-1,\ldots,Ns-n_{\max}$
which cannot be represented in terms of the independent localized magnons
of smallest area. 
Thus, it remains unclear 
whether these additional states 
contribute to the partition function in the thermodynamic limit.
In our numerical data 
(see Tables \ref{tab2} and \ref{tab3} below) 
we also find extra spin states 
for the diamond chain with $J_2=2J_1$, 
the square-kagom\'{e} lattice with $J_2=J_1$, 
the two-leg ladder with $J_2=2J_1$ 
and 
two kagom\'{e}-like chains 
(but not for the diamond chain with $J_2=3J_1$, 
the dimer-plaquette chain with $J_2=J_1$, $J_3=2J_1$,
the square-kagom\'{e} lattice with $J_2=2J_1$,  
the sawtooth chain
and
the two-leg ladder with $J_2=3J_1$).
For all these systems, 
except the square-kagom\'{e} lattice with $J_2=J_1$,   
the numbers of the extra spin states do not increase 
with increasing of the lattice sizes 
that indicates their irrelevance in the thermodynamic limit.
For the square-kagom\'{e} lattice with $J_2=J_1$,
as for the  checkerboard\cite{25} and the kagom\'{e}\cite{26} lattices, 
we find much more extra spin states,
however, 
our data restricted to finite systems 
are not sufficient to find reasonable tendencies for the thermodynamic limit. 
Furthermore, we have to discuss whether 
the independent localized magnon states 
are separated by a finite energy from higher-energy states.
We postpone a discussion of this issue till Sec.~\ref{4} 
where the results of the exact diagonalization for finite systems 
are presented.

We are interested in the contribution 
of the independent localized magnons of smallest area 
to the thermodynamics of the spin systems.
To find this contribution 
we must calculate 
the part of the partition function of the spin system,
$Z(T,h,N)=\sum_j\exp\left(-E_j(h,N)/kT\right)$
($j$ denotes all states of the system of $N$ spins $s$ on a lattice),
which comes from the independent localized magnons.
Denoting this quantity by $Z_{\rm{lm}}(T,h,N)$
we have
\begin{eqnarray}
Z_{\rm{lm}}(T,h,N)
=
\sum_{n=0}^{n_{\max}}
g_N(n)\exp\left(-\frac{E_n(h)}{kT}\right)
=
\exp\left(-\frac{E_{\rm{FM}}-hsN}{kT}\right)
\sum_{n=0}^{n_{\max}}
g_N(n)
\exp\left(\frac{\mu}{kT}n\right)
\label{03}
\end{eqnarray}
with $\mu=\epsilon_1-h$.

An important step for further calculation of $Z_{\rm{lm}}(T,h,N)$ (\ref{03})
is a mapping onto a hard-core object lattice gas.
Consider an auxiliary lattice 
in which each site can be occupied or not by a hard-core object 
(monomer, dimer, hexagon or square)
which corresponds to a localized magnon of smallest area
(see Figs. \ref{fig1}, \ref{fig2}).
Let us denote by ${\cal{N}}$ 
the number of sites of the auxiliary lattice;
the relation between $N$ and ${\cal{N}}$ 
for the considered spin systems is given in Table \ref{tab1}.
We note 
that $g_N(n)$ is simply the canonical partition function $Z(n,{\cal{N}})$ 
of $n$ hard-core objects 
placed on a lattice with ${\cal{N}}$ sites.
Furthermore,
$\Xi(T,\mu,{\cal{N}})
=\sum_{n}\exp(\mu n/kT)Z(n,{\cal{N}})$
is the grand canonical partition function 
of hard-core objects 
placed on a lattice with ${\cal{N}}$ sites
and $\mu$ is the chemical potential.
As a result,
according to Eq. (\ref{03}) 
we arrive at the basic relation between the thermodynamic quantities 
of independent localized magnons and hard-core models,
i.e. 
$Z_{\rm{lm}}(T,h,N)
=
\exp(-(E_{\rm{FM}}-hsN)/kT)\Xi(T,\mu,{\cal{N}})$.

The calculation of $\Xi(T,\mu,{\cal{N}})$ 
for a classical hard-core object lattice gas 
(usually in the thermodynamic limit), 
in general, 
is a nontrivial problem \cite{27}.
It may be convenient to perform such calculations 
using the occupation number representation.
Let us introduce the occupation number $n_i$ 
which takes two values, 0 and 1, 
depending whether the site $i$ of the auxiliary lattice 
is empty or occupied, $i=1,\ldots,{\cal{N}}$.
Then the grand canonical partition function of hard-core objects 
can be rewritten in the following way
\begin{eqnarray}
\Xi(T,\mu,{\cal{N}})
=
\sum_{n_1=0,1}\ldots\sum_{n_{\cal{N}}=0,1}
\exp\left(\frac{\mu}{kT}\sum_{i=1}^{{\cal{N}}}n_i\right)
R(n_1,\ldots,n_{\cal{N}}),
\label{04}
\end{eqnarray}
where the function $R(n_1,\ldots,n_{\cal{N}})\ne 1$
appears if the hard-core objects are extended (dimers, hexagons, squares) 
rather than simple monomers.
For example, 
for the sawtooth chain,
for which the corresponding hard-core objects are rigid dimers,
we have
$R(n_1,\ldots,n_{\cal{N}})=\prod_{i}(1-n_in_{i+1})$
and all the terms in (\ref{04})
which correspond to two adjacent sites of the auxiliary lattice 
being occupied have dropped out.

Using the relation between 
the independent localized magnons partition function 
$Z_{\rm{lm}}(T,h,N)$ 
and the grand canonical partition function 
of hard-core object lattice gas
$\Xi(T,\mu,{\cal{N}})$
we get the following result for 
the Helmholtz free energy (per site) of the independent localized magnons
\begin{eqnarray}
\frac{F_{\rm{lm}}(T,h,N)}{N}
=
\frac{E_{\rm{FM}}}{N}-hs
-kT\frac{\ln\Xi(T,\mu,{\cal{N}})}{N}.
\label{05}
\end{eqnarray}
The entropy 
$S_{\rm{lm}}(T,h,N)
=-\partial F_{\rm{lm}}(T,h,N)/\partial T$
and the specific heat 
$C_{\rm{lm}}(T,h,N)
=T\partial S_{\rm{lm}}(T,h,N)/\partial T$
follow immediately from Eq. (\ref{05}). 
We can also calculate 
the average number of hard-core objects
$\langle n\rangle=\sum_{i=1}^{{\cal{N}}}\langle n_i\rangle
=kT\partial
\ln\Xi(T,\mu,{\cal{N}})/\partial\mu$
which yields the magnetization
$\langle S^z\rangle_{\rm{lm}}=sN-\langle n\rangle$.

Next, we turn to the specific lattice gases 
which are related to the considered spin systems.
Although the results for various lattice gases 
are not new and can be found in the literature \cite{27}
we present some of them here for easy references 
in view of further discussions for the spin systems.

\subsection{Monomers}
\label{3.1}

\begin{figure}
\begin{center}
\includegraphics[clip=on,width=58mm,angle=0]{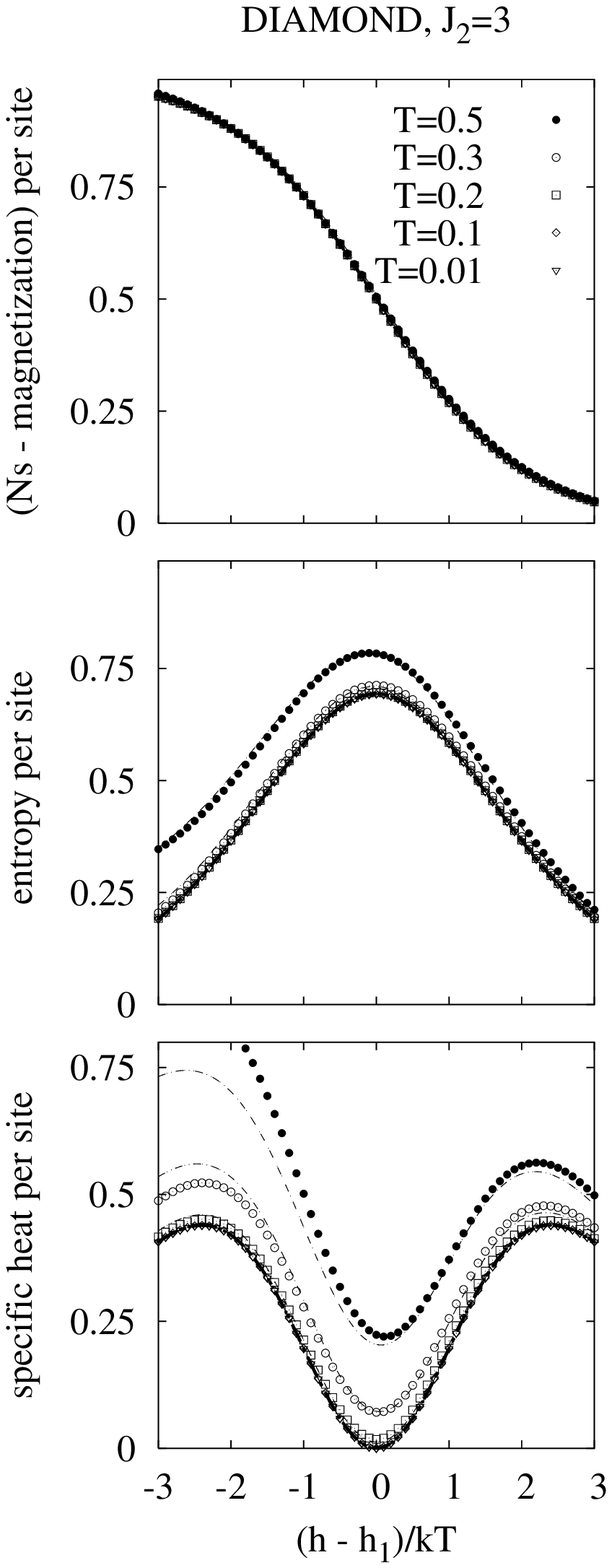}
\hspace{-2mm}
\includegraphics[clip=on,width=58mm,angle=0]{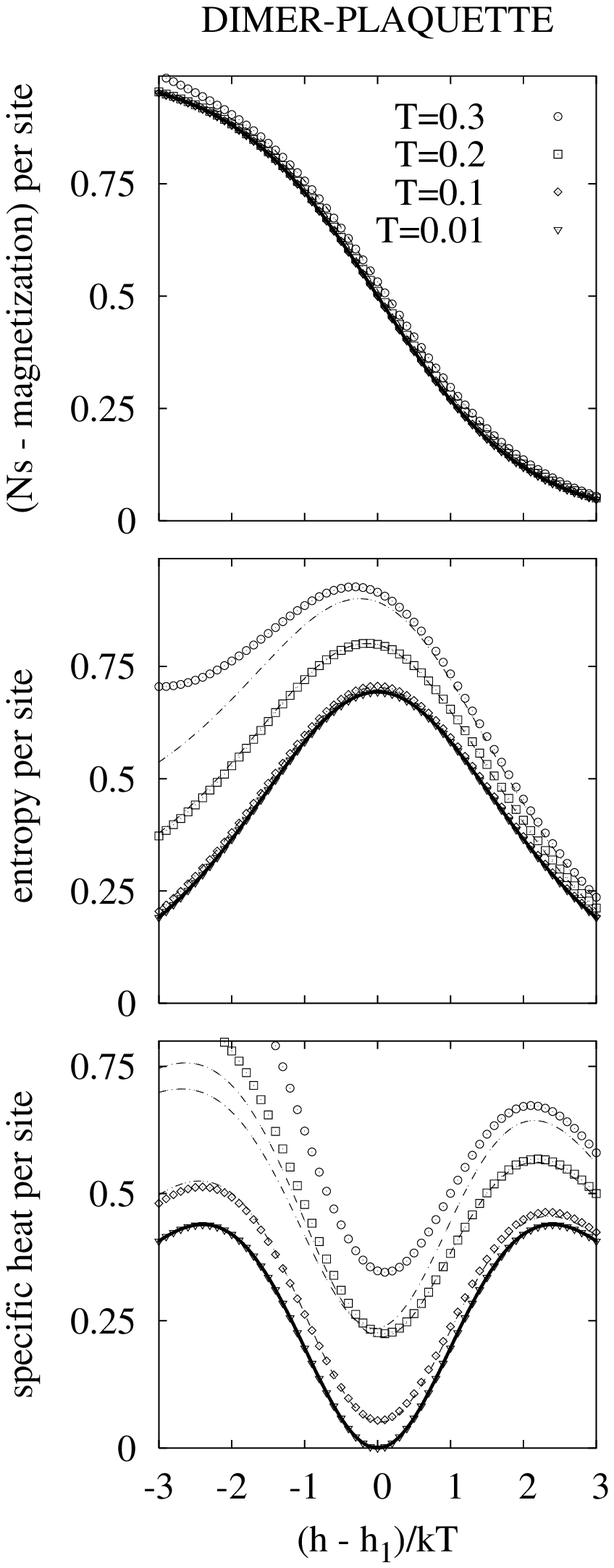}
\hspace{-2mm}
\includegraphics[clip=on,width=58mm,angle=0]{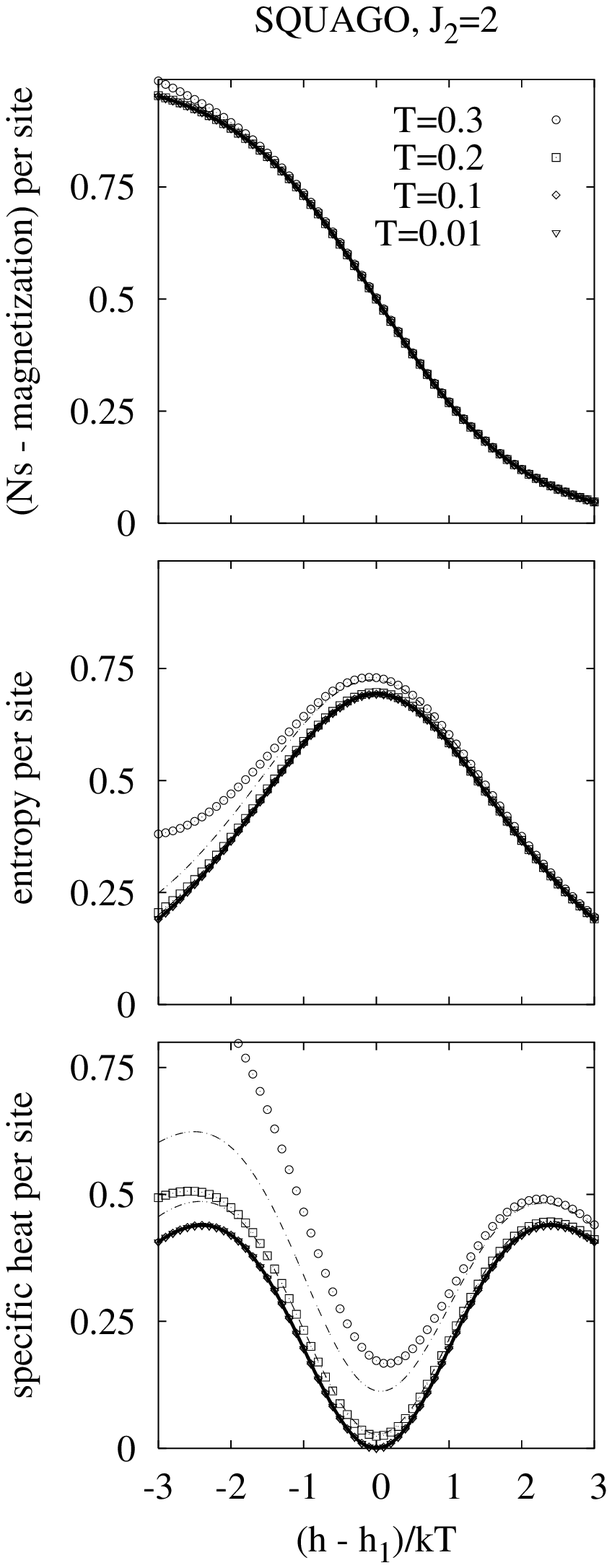}
\end{center}
\caption[]
{The universal dependences of
$(sN-\langle S^z\rangle_{\rm{lm}})/{\cal{N}}$ (\ref{09}),
$S_{\rm{lm}}/k{\cal{N}}$ (\ref{07})
and
$C_{\rm{lm}}/k{\cal{N}}$ (\ref{08}) 
(monomer universality class)
together with the corresponding results 
for finite spin systems (diamond chain with $N=18$, $J_1=1$, $J_2=3$ [left column], 
dimer-plaquette chain with $N=20$, $J_1=J_2=1$, $J_3=2$ [middle column], 
square-kagom\'{e} lattice with $N=18$, $J_1=1$, $J_2=2$ [right column]). 
The universal dependences are shown by solid lines,
the exact diagonalization data are shown by symbols. 
Note that
the exact diagonalization data for the lowest temperature 
coincide with the universal dependences for monomers.
We also show by thin broken lines the analytical results 
which follow from (\ref{10}) 
with $U=1.57,\;0.58,\;1.34$ 
for the diamond chain, dimer-plaquette chain and square-kagom\'{e} lattice,
respectively,
which are in good agreement with numerical data for higher temperatures. 
\label{fig3}}
\end{figure}

\begin{figure}
\begin{center}
\includegraphics[clip=on,width=58mm,angle=0]{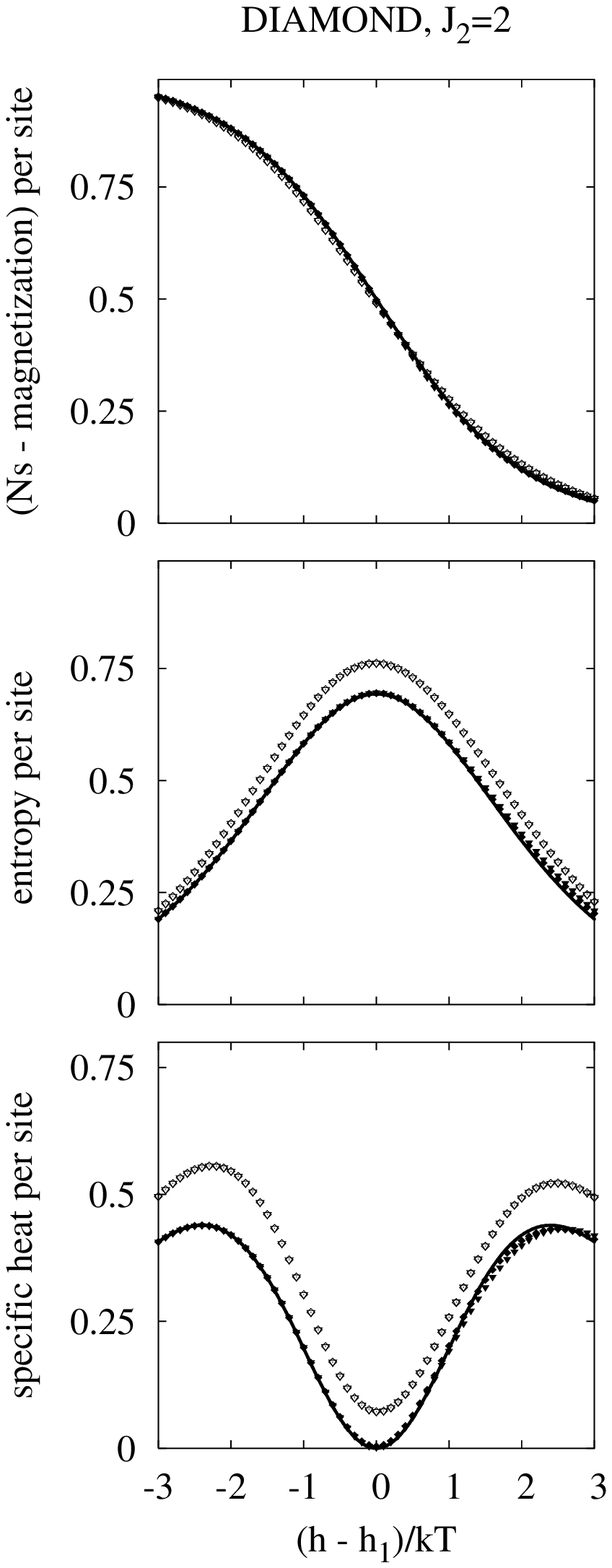}
\hspace{0mm}
\includegraphics[clip=on,width=58mm,angle=0]{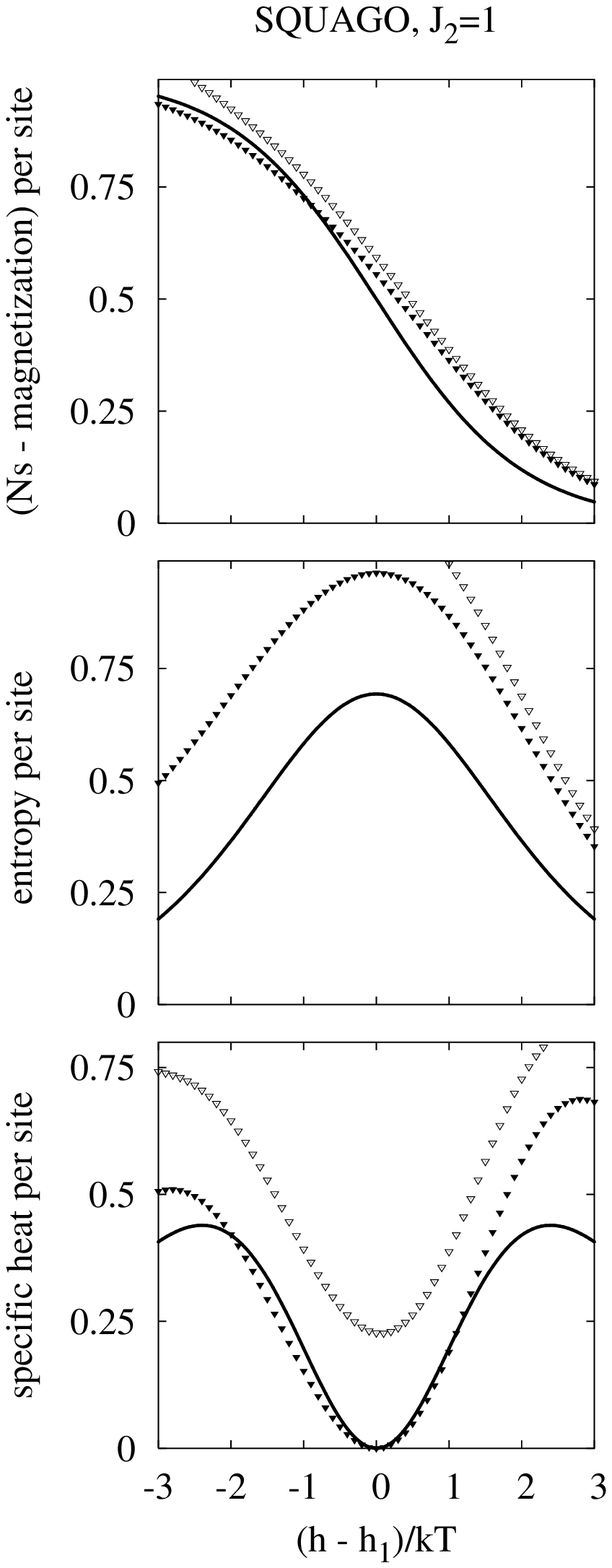}
\end{center}
\caption[]
{The universal dependences for monomers 
(solid lines) 
of
$(sN-\langle S^z\rangle_{\rm{lm}})/{\cal{N}}$,
$S_{\rm{lm}}/k{\cal{N}}$
and
$C_{\rm{lm}}/k{\cal{N}}$
together with the corresponding results for 
finite spin systems
belonging to the monomer universality class
but which have extra (not hard-monomer) spin states:
the diamond chain with $N=18$, $J_1=1$, $J_2=2$ 
(left column),
the square-kagom\'{e} lattice with $N=18$,  $J_1=J_2=1$ 
(right column).
The numerical results  for spin systems are shown for  temperatures 
$kT=0.01$ (filled symbols) 
and
$kT=0.1$ (empty symbols).
\label{fig4}}
\end{figure}

The independent localized magnon states 
in the diamond and dimer-plaquette chains 
as well as in the square-kagom\'{e} lattice 
can be mapped onto a lattice gas of monomers
(see panels a and b in Fig. \ref{fig1}, panel a in Fig. \ref{fig2}
and Table \ref{tab1}). 
The calculation of the thermodynamic quantities 
for the lattice gas of monomers is simple. 
We have
$Z(n,{\cal{N}})={\cal{C}}_{{\cal{N}}}^n
={\cal{N}}!/(n!({\cal{N}}-n)!)$,
$\Xi(T,\mu,{\cal{N}})
=\sum_{n=0}^{{\cal{N}}}{\cal{C}}_{{\cal{N}}}^n\exp(\mu n/kT)
=(\exp(\mu/kT) +1)^{{\cal{N}}}$
or, using Eq. (\ref{04}),
$\Xi(T,\mu,{\cal{N}})
=(\sum_{n=0,1}\exp(\mu n/kT))^{\cal{N}}
=(1+\exp(\mu/kT))^{\cal{N}}$.
The Helmholtz free energy of the gas of monomers reads 
\begin{eqnarray}
\frac{F_{\rm{lm}}(T,h,N)}{N}
=
\frac{E_{\rm{FM}}}{N}-hs
-\frac{{\cal{N}}}{N}
kT\ln\left(1+\exp\frac{\mu}{kT}\right)
\label{06}
\end{eqnarray}
and therefore 
\begin{eqnarray}
\frac{S_{\rm{lm}}(T,h,N)}{kN}
=
\frac{{\cal{N}}}{N}
\left(\ln\left(1+\exp x\right)-\frac{x\exp x}{1+\exp x}
\right),
\label{07}
\end{eqnarray}
\begin{eqnarray}
\frac{C_{\rm{lm}}(T,h,N)}{kN}
=
\frac{{\cal{N}}}{N}
\frac{\left(\frac{x}{2}\right)^2}{\cosh^2\frac{x}{2}},
\label{08}
\end{eqnarray}
\begin{eqnarray}
\frac{\langle S^z\rangle_{\rm{lm}}}{sN}
=1-\frac{1}{s}\frac{{\cal{N}}}{N}\frac{\exp x}{1+\exp x}
\label{09}
\end{eqnarray}
with $x=\mu/kT$.
Eqs. (\ref{07}), (\ref{08}) are symmetric with respect to $x\to -x$.
We note 
that at saturation $\mu=0$ and Eq. (\ref{07}) yields
$S_{\rm{lm}}(T,h_1,N)/kN
=({\cal{N}}/N)\ln 2$.
That is the value of the residual entropy per site 
obtained earlier for the diamond and dimer-plaquette chains\cite{11}.
The specific heat $C_{\rm{lm}}(T,h,N)/k{\cal{N}}$, Eq.~(\ref{08}), 
exhibits two symmetric maxima of height 
$C_{\max} \approx 0.43922884$ 
at $x\approx \pm 2.39935728$.
The universal dependences 
of $(Ns-\langle S^z\rangle_{\rm{lm}})/{\cal{N}}$, Eq.~(\ref{09}),
of $S_{\rm{lm}}/k{\cal{N}}$, Eq.~(\ref{07}), and
of $C_{\rm{lm}}/k{\cal{N}}$, Eq.~(\ref{08}),
are shown in Figs.~\ref{fig3}, \ref{fig4} (for a more detailed 
discussion of these Figures, see Sec.~\ref{4}). 

We may consider an improvement of a lattice gas description 
of low-lying energy levels of spin systems 
taking into account higher-energy states 
which are separated from the hard-core object states by a gap.
According to Eq. (\ref{04}) 
within a lattice gas model there are only two states at each lattice site.
Namely, 
the site may be either empty ($n_i=0$) 
and such a state occurs with the probability $\propto 1$,
or 
the site may be occupied by monomer ($n_i=1$) 
and such a state occurs with the probability $\propto \exp(\mu/kT)$.
We may increase the number of states at each sites 
introducing an additional higher-energy state of monomers 
which occurs with the probability $\propto \exp((\mu-U)/kT)$
and $U$ is large positive on-site energy.
Obviously, 
if $U\to\infty$ the higher-energy states are irrelevant, 
but for finite $U$ they become relevant as temperature grows.
In the spin picture,
the second state corresponds to a higher-energy state 
with a wave function located in the area 
which is mapped onto one site of the auxiliary lattice.
Our calculations confirm that the made assumption 
really yields an improvement for higher temperatures
(see the corresponding panels in Figs.~\ref{fig3}, \ref{fig7}, \ref{fig8}).
The grand partition function of the modified lattice-gas model 
can be easily calculated
\begin{eqnarray}
\Xi(T,\mu,{\cal{N}})
=\left(1+\exp\frac{\mu}{kT}+\exp\frac{\mu-U}{kT}\right)^{{\cal{N}}}.
\label{10}
\end{eqnarray}
From Eq. (\ref{10}) we find the entropy, the specific heat, and the 
magnetization using usual thermodynamic  relations.
The corresponding results 
are illustrated
in Figs. \ref{fig3}, \ref{fig7}, \ref{fig8}  
assuming values for $U$ 
related to the energy separation between the 
localized magnon states  and higher-energy states for finite 
spin lattices, see Sec.~\ref{4}.
The improved approximation agrees with exact diagonalization results 
up to higher temperatures in comparison with the monomer lattice gas 
predictions (\ref{06}) -- (\ref{09}).

\subsection{One-dimensional dimers}
\label{3.2}

\begin{figure}
\begin{center}
\includegraphics[clip=on,width=58mm,angle=0]{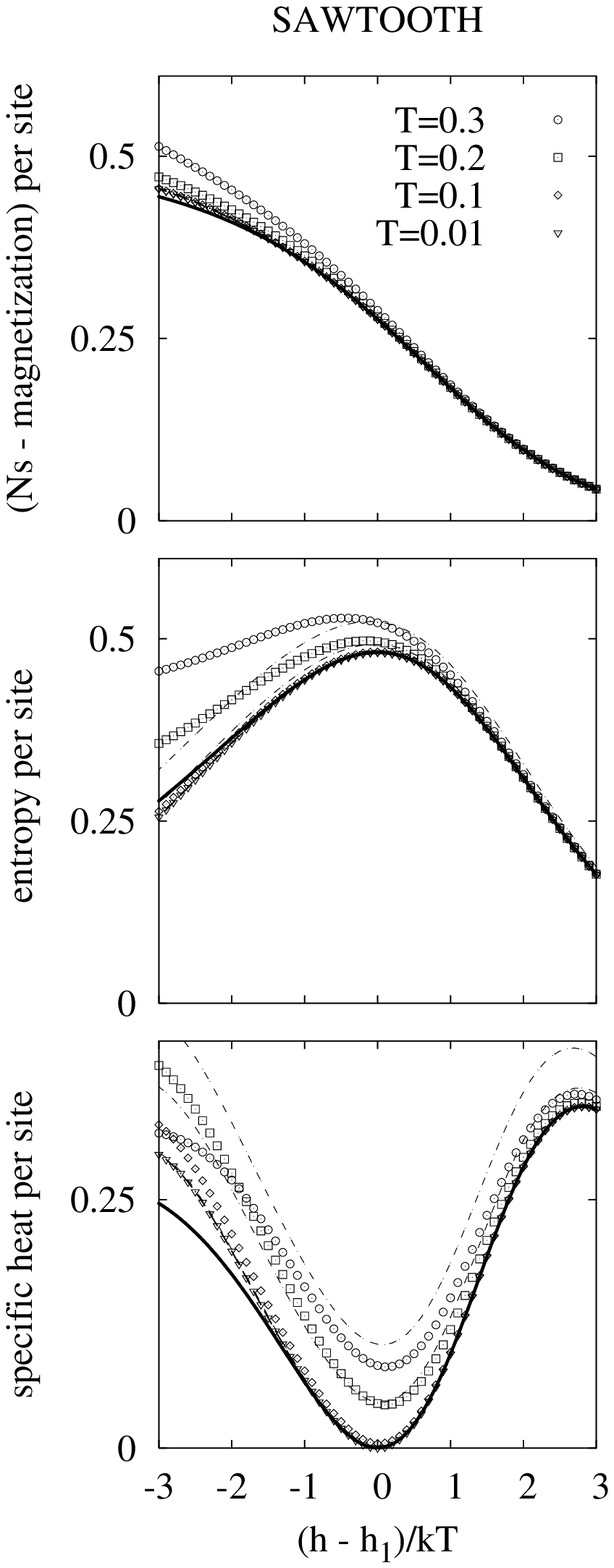}
\hspace{0mm}
\includegraphics[clip=on,width=58mm,angle=0]{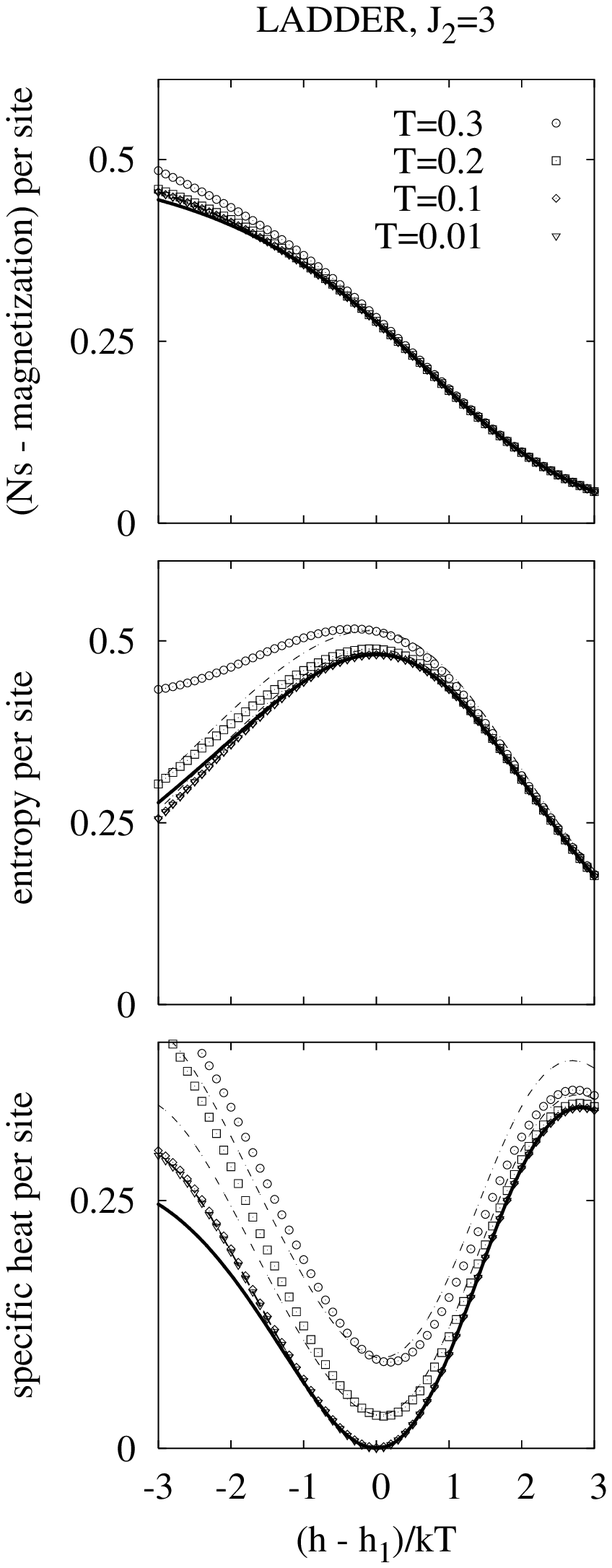}
\end{center}
\caption[]
{The universal dependences of
$(sN-\langle S^z\rangle_{\rm{lm}})/{\cal{N}}$, 
$S_{\rm{lm}}/k{\cal{N}}$ 
and
$C_{\rm{lm}}/k{\cal{N}}$ 
(one-dimensional dimer universality class)
together with the corresponding results 
for finite spin systems (sawtooth chain with $N=20$, $J_1=1$, $J_2=2$ [left column] 
and 
two-leg ladder with $N=20$, $J=1$, $J_2=3$ [right column]).
The universal dependences 
which follow from (\ref{11})
are shown by thick solid (${\cal{N}}\to\infty$) and dashed lines (${\cal{N}}=10$),
the exact diagonalization data are shown by symbols. Note that
the exact diagonalization data for the lowest temperature 
coincide with the universal dependences for dimers.
We also show by thin broken lines the analytical results 
which follow from (\ref{18}) for ${\cal{N}}=10$ 
with $U=1.00$ and $1.10$                                             
for the sawtooth chain and two-leg ladder,
respectively,
which are in good agreement with numerical data for higher temperatures.
\label{fig5}}
\end{figure}

\begin{figure}
\begin{center}
\includegraphics[clip=on,width=58mm,angle=0]{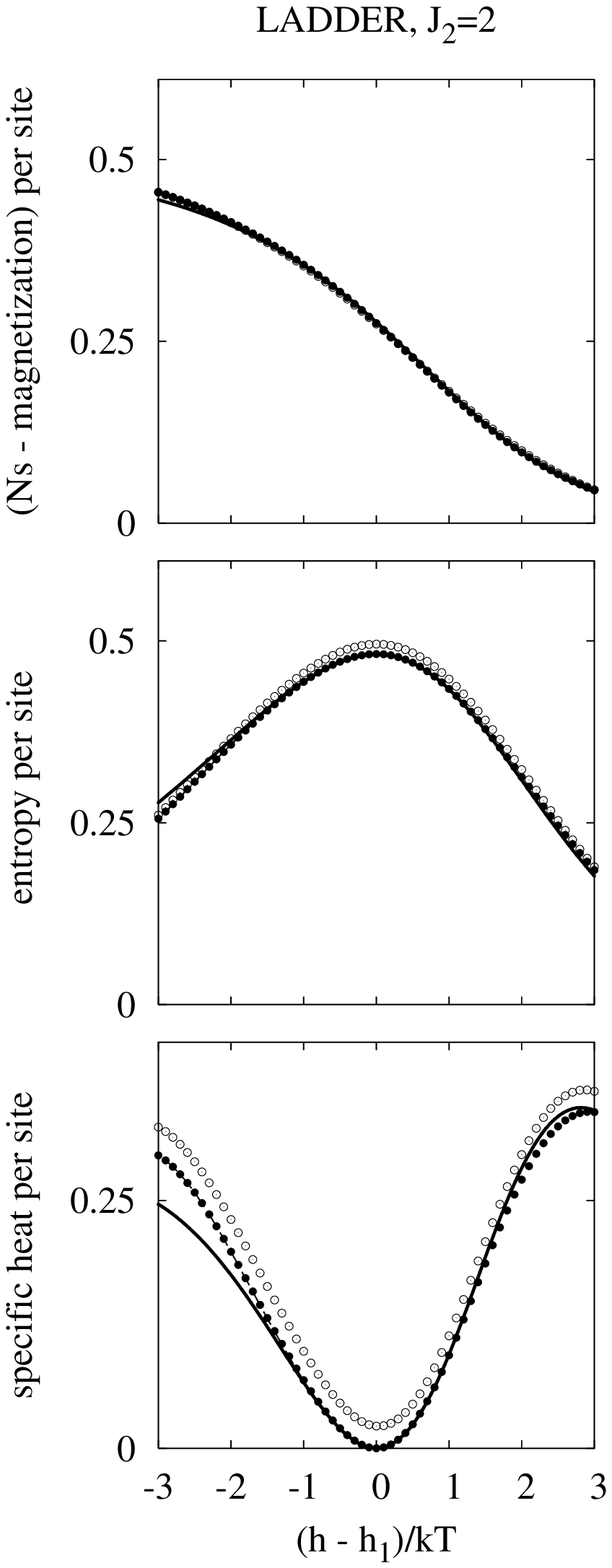}
\hspace{-2mm}
\includegraphics[clip=on,width=58mm,angle=0]{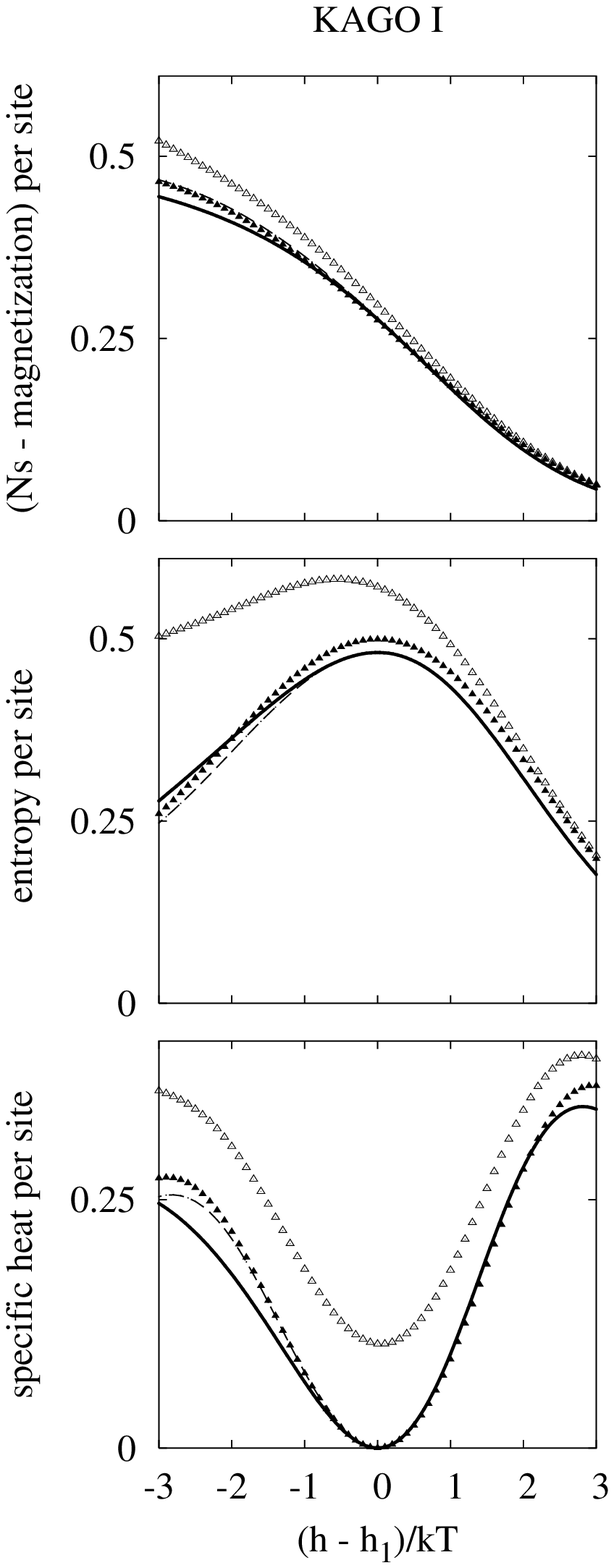}
\hspace{-2mm}
\includegraphics[clip=on,width=58mm,angle=0]{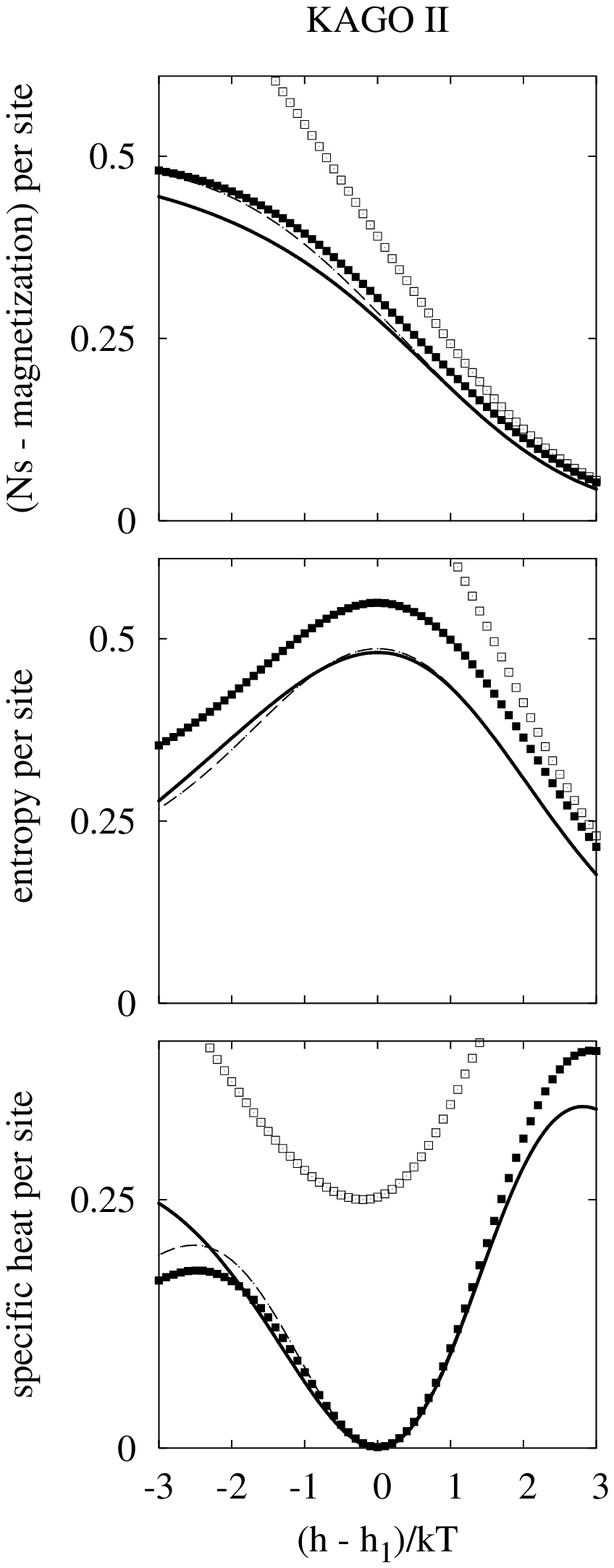}
\end{center}
\caption[]
{The universal dependences of
$(sN-\langle S^z\rangle_{\rm{lm}})/{\cal{N}}$,
$S_{\rm{lm}}/k{\cal{N}}$
and
$C_{\rm{lm}}/k{\cal{N}}$
for one-dimensional dimers 
(${\cal N} \to \infty$ -- thick solid lines; finite $\cal N$ -- dashed lines) 
together with the corresponding results for 
finite spin systems
belonging to the one-dimensional dimer universality class
but which have extra (not hard-dimer) spin states:
the two-leg ladder with $N=20$ (corresponds to ${\cal N}=10$), $J_1=1$, $J_2=2$ 
(left column), 
the kagom\'{e}-like chain shown in Fig. \ref{fig1}e with $N=18$ (corresponds to ${\cal N}=6$), $J=1$ 
(middle column)
and
the kagom\'{e}-like chain shown in Fig. \ref{fig1}f with $N=20$ (corresponds to ${\cal N}=4$), $J_1=1$ 
(right column).
The numerical results  for spin systems are shown for  temperatures 
$kT=0.01$ (filled symbols) 
and
$kT=0.1$ (empty symbols).

\label{fig6}}
\end{figure}

The independent localized magnons 
in the sawtooth chain, the two-leg ladder and two kagom\'{e}-like chains 
can be mapped onto a one-dimensional lattice gas of rigid dimers
(see panels c, d, e, f in Fig. \ref{fig1}
and Table \ref{tab1}). 
The grand partition function $\Xi(T,\mu,{\cal{N}})$ (\ref{04})
of rigid dimers on a chain of ${\cal{N}}$ sites 
with periodic boundary conditions
can be calculated using the transfer-matrix method.
The resulting expression for $\Xi(T,\mu,{\cal{N}})$ reads 
\begin{eqnarray}
\Xi(T,\mu,{\cal{N}})
=\lambda_1^{{\cal{N}}}+\lambda_2^{{\cal{N}}},
\;\;\;
\lambda_{1,2}
=
\frac{1}{2}\pm\sqrt{\frac{1}{4}+\exp\frac{\mu}{kT}}.
\label{11}
\end{eqnarray}
In the thermodynamic limit ${\cal{N}}\to\infty$
only the larger eigenvalue of the transfer matrix plays role in (\ref{11}) 
and we have
\begin{eqnarray}
\frac{F_{\rm{lm}}(T,h,N)}{N}
=
\frac{E_{\rm{FM}}}{N}-hs
-\frac{{\cal{N}}}{N}kT
\ln\left(
\frac{1}{2}
+\sqrt{\frac{1}{4}+\exp\frac{\mu}{kT}}
\right).
\label{12}
\end{eqnarray}
For the entropy, specific heat, and magnetization we find from Eq. (\ref{12})
\begin{eqnarray}
\frac{S_{\rm{lm}}(T,h,N)}{kN}
=
\frac{{\cal{N}}}{N}
\left(
\ln\left(\frac{1}{2}+\sqrt{\frac{1}{4}+\exp x}\right)
-x\left(\frac{1}{2}-\frac{1}{4\sqrt{\frac{1}{4}+\exp x}}\right)
\right),
\label{13}
\end{eqnarray}
\begin{eqnarray}
\frac{C_{\rm{lm}}(T,h,N)}{kN}
=
\frac{{\cal{N}}}{N}
\frac{x^2\exp x}{8\left(\frac{1}{4}+\exp x\right)^\frac{3}{2}},
\label{14}
\end{eqnarray}
\begin{eqnarray}
\frac{\langle S^z\rangle_{\rm{lm}}}{sN}
=1-\frac{1}{s}\frac{{\cal{N}}}{N}
\left(\frac{1}{2}-\frac{1}{4\sqrt{\frac{1}{4}+\exp x}}\right),
\label{15}
\end{eqnarray}
respectively.
Note that Eqs. (\ref{13}), (\ref{14}) 
are not symmetric with respect to $x\to -x$.
Eqs. (\ref{13}), (\ref{14}), (\ref{15}) 
were examined in Refs. \cite{12,13,08}
in the context of the sawtooth chain.
We also note 
that at saturation, i.e. at $\mu=0$, 
Eq. (\ref{13}) gives the values of the residual entropy per site 
for the kagom\'{e}-like chains and the two-leg ladder,
$S_{\rm{lm}}(T,h_1,N)/kN
=({\cal{N}}/N)\ln((1+\sqrt{5})/2)$ \cite{11}.
The universal dependences  
$(Ns-\langle S^z\rangle_{\rm{lm}})/{\cal{N}}$, Eq.~(\ref{15}),
$S_{\rm{lm}}/k{\cal{N}}$, Eq.~(\ref{13}),
and
$C_{\rm{lm}}/k{\cal{N}}$, Eq.~(\ref{14}),
are shown in Figs.~\ref{fig5} and \ref{fig6} 
(for a more detailed 
discussion of these Figures, see Sec.~\ref{4}).
The specific heat $C_{\rm{lm}}(T,h,N)/k{\cal{N}}$ exhibits two maxima of 
heights 
$C^{{\rm{right}}}_{\max}\approx 0.34394234$ 
and 
$C^{{\rm{left}}}_{\max}\approx 0.26887020$ 
at 
$(h-h_1)/kT \approx 2.81588498$ 
and 
$(h-h_1)/kT \approx -4.05258891$, 
respectively. 

Again we may consider an improvement of a lattice gas description
of the low-lying energy levels of spin systems.
One way to consider higher-energy states 
was suggested by M.~E.~Zhitomirsky and H.~Tsunetsugu \cite{08}.
These authors treat the  
factor $1-n_in_{i+1}$ in (\ref{04}),
as the $V\to\infty$ limit 
of a generalized factor
$\exp(-(V/kT) n_in_{i+1})
=1+(\exp(-V/kT)-1)n_in_{i+1}$.
We may relax the hard-core restriction assuming 
the intersite interaction $V$ to be large but finite
and approximately equal to the 
 energy separation between the 
localized magnon states  and higher-energy states
(in Ref. \cite{08} the energy gap was also estimated 
using variational calculations).
As a result,
we arrive at a one-dimensional lattice gas 
with a finite nearest-neighbor repulsion.
The grand partition function of such a gas is calculated 
by the transfer-matrix method and reads
\begin{eqnarray}
\Xi(T,\mu,{\cal{N}})
&=&\lambda_1^{{\cal{N}}}+\lambda_2^{{\cal{N}}},
\nonumber\\
\lambda_{1,2}&=&\frac{1}{2}+\frac{1}{2}\exp\frac{\mu-V}{kT}
\pm
\sqrt{\frac{1}{4}+\exp\frac{\mu}{kT}-\frac{1}{2}\exp\frac{\mu-V}{kT}
+\frac{1}{4}\exp\frac{2(\mu-V)}{kT}}.
\label{16}
\end{eqnarray}
The thermodynamic quantities which follow from Eq. (\ref{16}) 
are in a good agreement with exact diagonalization data for the sawtooth chain 
up to higher temperatures 
in comparison with the predictions in the $V\to\infty$ limit \cite{08}. 
In the spin picture a finite $V$  
corresponds to higher-energy states 
with wave functions located on the areas
which are mapped onto two adjacent sites of the auxiliary one-dimensional lattice. 
In addition, we can consider a similar improvement 
as for the monomer problem, namely we can assume additional higher-energy states of dimers  
which occur with the probability 
$\propto\exp((\mu-U)/kT)$, 
where $U$ is positive and large. That means that more than one dimer is allowed on the same site, 
but that cost additional energy $U$.    
In the spin picture,
we take into account  a higher-energy state 
with a wave function located in the area
which is mapped onto one site of the auxiliary one-dimensional lattice. 
Thus, combining 
the assumption about higher-energy state of dimer at each site 
and
the weakening of the hard-core restriction 
we arrive at the following formula for the grand partition function 
\begin{eqnarray}
\Xi(T,h,{\cal{N}})
=\lambda_1^{{\cal{N}}}+\lambda_2^{{\cal{N}}}+\lambda_3^{{\cal{N}}}, \hspace{10cm}
\nonumber\\
\lambda^3
-\left(1+\exp\frac{\mu-V}{kT}\right)\lambda^2
-\exp\frac{\mu}{kT}\left(1+\exp\left(-\frac{U}{kT}\right)-\exp\left(-\frac{V}{kT}\right)\right)\lambda
+\exp\frac{2\mu-U-V}{kT}
=0;
\label{17}
\end{eqnarray}
in the thermodynamic limit ${\cal{N}}\to\infty$ 
only the largest root of the qubic equation enters the thermodynamic quantities. 
In the limit $U\to\infty$ 
Eq. (\ref{17}) transforms into Eq. (\ref{16}).
In what follows we consider only the other limit $V\to\infty$ 
when Eq. (\ref{17}) becomes
\begin{eqnarray}
\Xi(T,\mu,{\cal{N}})
&=&\lambda_1^{{\cal{N}}}+\lambda_2^{{\cal{N}}},
\nonumber\\
\lambda_{1,2}&=&\frac{1}{2}
\pm
\sqrt{\frac{1}{4}+\exp\frac{\mu}{kT}+\exp\frac{\mu-U}{kT}}.
\label{18}
\end{eqnarray}
From Eq. (\ref{18}) we find in a usual way
the entropy, the specific heat, and the magnetization, which are shown 
in Figs. \ref{fig5}, \ref{fig7}, \ref{fig8}  
assuming values for $U$ 
again obtained by inspection of the energy spectrum of finite 
spin lattices (for more details see Sec.~\ref{4}).

\subsection{Hexagons on a triangular lattice 
           and squares on a square lattice}
\label{3.3}

For completeness we give here some remarks 
on the two-dimensional lattice gases of hard-core objects 
which emerge in the context of localized magnons 
in some two-dimensional frustrated quantum antiferromagnets. 
Such lattice gases are much more complicated subject 
for rigorous analytical or numerical studies.

For the kagom\'{e} and star lattices 
the independent localized magnons of smallest area 
can be put in correspondence 
to the hard  hexagons on a triangular lattice 
(see panels b and c in Fig. \ref{fig2}
and Table \ref{tab1}).
The grand canonical partition function 
for the hard hexagons on triangular lattice,
$\Xi(T,\mu,{\cal{N}})$,
was found by R.~J.~Baxter 
(hard-hexagon problem, see Ref.~\cite{27}).
The parametric dependence 
of $(\Xi(T,\mu,{\cal{N}}))^{1/{\cal{N}}}$ 
on the activity $z=\exp(\mu/kT)$
in the thermodynamic limit ${\cal{N}}\to\infty$
can be found in Ref.~\cite{27}.
The peculiarity of the hard-hexagon model 
is the existence of a phase transition 
at a critical value of activity  
$z_c=(11+5\sqrt{5})/2\approx 11.09016994$
($x_c=\ln z_c\approx 2.40605913$).
The low-activity phase is uniform (liquid)
with identical average occupation numbers of hexagons on each of three triangular sublattices,
$n_1=n_2=n_3$,
whereas the high-activity phase is nonuniform (solid)
with $n_1>n_2=n_3$. 
The value of the entropy per site at $z=1$ is $0.33324272 k$;
this number gives the residual entropy per site due to localized magnon states 
of smallest area
for the kagom\'{e} and star lattices  \cite{11,13,08}.
The hard-hexagon problem 
in the context of the low-temperature magnetothermodynamics of the kagom\'{e} lattice 
in the vicinity of the saturation field
was discussed in detail in Refs. \cite{13,08}.

For the checkerboard lattice 
the independent localized magnons of smallest area 
can be put in correspondence to the hard-core squares on a square lattice, where
the size of the hard squares 
corresponds to 
the nearest-neighbor 
and 
the next-nearest-neighbor exclusions
(see panel d in Fig. \ref{fig2}
and Table \ref{tab1}).
We are not aware for an exact analytical solution of such a model, but approximative results are available\cite{28}. 
In the context of the low-temperature magnetothermodynamics of the checkerboard 
lattice 
in the vicinity of the saturation field
the hard-square problem has been discussed recently in Ref. \cite{29}.
The value of the entropy per site at $z=1$ 
obtained applying a classical Monte Carlo method 
is about $0.2946 k$ \cite{29} 
(a direct calculation for periodic $8\times 8$ ($8\times 10$) lattice
yields about $0.2949 k$ ($0.2948 k$) \cite{30}).

\section{Exact diagonalization versus hard-core object description}
\label{4}

After presenting explicit  expressions for the thermodynamic quantities by
using the correspondence between the localized magnon states of the quantum
spin system and a classical hard-core description we now test our analytical
results by comparison with finite-lattice numerical results of the full spin-1/2 
system. Note that the classical equations for the monomer problem, see
Sec.~\ref{3.1}, do not depend on the size of the system, whereas the
corresponding classical equations for the dimer problem, see Sec.~\ref{3.2}, are
size dependent.
For the number of degenerate states presented in
Tables~\ref{tab2} and \ref{tab3} we could consider systems of up to $N=30$ spins, 
since we can restrict the calculation of sectors with high total $S^z$.
For the thermodynamics we used the data of full diagonalization of spin systems of 
either $N=18$ or $N=20$.    
For the estimation of the temperature and field ranges, 
in which the hard-core description is valid, and also for the estimation 
of the  on-site energy parameter $U$, see Secs.~\ref{3.1} and \ref{3.2}, it is 
useful  to find a measure for the thermodynamically relevant 
energy separation $\Delta$ between the localized magnon states and the other 
eigenstates of the spin system. For that we have calculated the integrated
low-energy 
density of states  at saturation field and define $\Delta$ as 
that energy value above the localized magnon 
ground-state energy,  where the contribution of the higher-energy states 
to the  integrated  density of states becomes as large as the 
contribution of the localized-magnon states.  

In Figs.~\ref{fig3} and \ref{fig4}
we present the results for  the magnetization $(Ns-\langle S^z\rangle)/{\cal{N}}$,
the entropy $S/k{\cal{N}}$,
the specific heat $C/k{\cal{N}}$ in dependence on $(h-h_1)/kT$
for spin systems belonging 
to the monomer universality class, 
i.e. for the diamond chain, the dimer-plaquette chain and the 
square-kagom\'{e} lattice.
Note that the universal formulas (\ref{06}) -- (\ref{09}) 
depend only on $(h-h_1)/kT$, i.e. they are identical for all temperatures 
$kT=0.01,\;0.1,\;0.2,\;0.3,\;0.5$ 
considered in Figs.~\ref{fig3} and \ref{fig4}.

Fig.~\ref{fig3} corresponds to spin systems, for which
the number of degenerate spin states at the saturation 
 ${\cal{W}}_{\rm{sp}}$ equals 
the number of states of the corresponding hard-core object lattice gas ${\cal{W}}$,
see Table \ref{tab2}.
Furthermore, these spin systems 
exhibit a quite large energy separation $\Delta$.
We find 
$\Delta \approx 1.5$ for the diamond chain 
($J_1=1$, $J_2=3$, $N=18$),
$\Delta \approx 0.6$ for the dimer-plaquette chain 
($J_1=J_2=1$, $J_3=2$, $N=20$),
and
$\Delta \approx 1.5$ for the square-kagom\'{e} lattice 
($J_1=1$, $J_2=2$, $N=18$).
Obviously, in Fig.~\ref{fig3} the curves for the spin systems
at temperatures up to $kT\approx 0.1 \approx 0.1 \Delta$ 
are almost identical with the universal curves.
But also for $kT > 0.1$ the qualitative behavior
of the entropy, the specific heat, and the magnetization
is quite well modeled by the universal formulas (\ref{07}), (\ref{08}), (\ref{09}).
The modified lattice-gas model, see Eq.~(\ref{10}), 
with parameters $U$ related to $\Delta$ indeed leads to 
an improved quantitative agreement  with the results 
for the spin systems at higher temperatures.     

Fig.~\ref{fig4} corresponds to spin systems, for which
the number of degenerate spin states at the saturation field 
 ${\cal{W}}_{\rm{sp}}$ is larger
than the number of states of the corresponding hard-core object lattice gas ${\cal{W}}$,
see Table \ref{tab2}. 
However, for the diamond chain with $J_2=2J_1$, we have only one extra spin 
state, ${\cal{W}}_{\rm{sp}}={\cal{W}}+1$, which 
becomes irrelevant for larger $N$.
By contrast,
there is a noticeable disagreement 
between ${\cal{W}}_{\rm{sp}}$ and ${\cal{W}}$
for the square-kagom\'{e} lattice with $J_2=J_1$
(Table \ref{tab2}).
Similarly to the kagom\'{e}\cite{13,08,26} and the checkerboard\cite{25} lattices,
for the square-kagom\'{e} lattice with $J_2=J_1$
we have extra spin states  
which are not covered 
by the hard-core description illustrated in Sec.~\ref{3.1}. 
The number of these extra spin states depends on system size $N$.
From our numerical data for 
$N=18,\;24,\;30$  
we are not able to conclude whether the localized magnon states of 
smallest area dominate 
the low-temperature thermodynamics in the thermodynamic limit.
Moreover, 
for these systems with extra spin states we estimate  significantly lower 
separations $\Delta$ of 
higher-energy states, namely 
$\Delta \approx 0.3$
for the diamond chain ($J_1=1$, $J_2=2$, $N=18$) and
 $\Delta \approx 0.2$ for
the square-kagom\'{e} lattice ($J_1=1$, $J_2=1$, $N=18$).
As a consequence 
we may find (see Fig.~\ref{fig4})
a perfect agreement between the spin and the hard-core data only for very low temperatures, but 
realize a quantitative deviation  already for $kT\approx 0.1$.
For the square-kagom\'{e} lattice with $J_1=1$, $J_2=1$ the larger number of extra spin states leads to  a
more pronounced deviation from the universal formulas. 
Nevertheless, the numerical data illustrate that 
the localized magnon states of 
smallest area contribute substantially to the partition function  thus leading  
to a low-temperature behavior which is at least qualitatively 
in agreement with the universal behavior 
given by Eqs.~(\ref{07}), (\ref{08}), and (\ref{09}).  
\begin{table}
\begin{center}
\caption
{Number of degenerate states: lattice gas of monomers and spin systems.
For a gas of monomers on ${\cal{N}}$ sites 
we have
${\cal{W}}=\Xi(T,\mu=0,{\cal{N}})=2^{{\cal{N}}}$.
\label{tab2}}
\vspace{5mm}
\begin{tabular}
{|c||c|c|c|c|c|c|} \hline
${\cal{N}}$ & ${\cal{W}}$ & diamond   & diamond    
                 & dimer-plaquette & square-kagom\'{e} & square-kagom\'{e} \\ 
            &             & ($J_2=2J_1$) & ($J_2=3J_1$)
                 & ($J_3=2J_1$)       & ($J_2=J_1$)    & ($J_2=2J_1$) \\ \hline \hline
 3  &    8 &    9 &    8 &   8 &   18  &  8 \\ \hline
 4  &   16 &   17 &   16 &  16 &   42  & 16 \\ \hline
 5  &   32 &   33 &   32 &  32 &   52  & 32 \\ \hline
 6  &   64 &   65 &   64 &  64 &    -  &  - \\ \hline
 7  &  128 &  129 &  128 &   - &    -  &  - \\ \hline
 8  &  256 &  257 &  256 &   - &    -  &  - \\ \hline
\end{tabular}
\end{center}
\end{table}
\begin{table}
\begin{center}
\caption
{Number of  degenerate states: one-dimensional hard-dimer lattice gas and spin systems.
For one-dimensional dimer gas on ${\cal{N}}$ sites we have
${\cal{W}}=\Xi(T,\mu=0,{\cal{N}})
=\lambda_1(\mu=0)^{\cal{N}}+\lambda_2(\mu=0)^{\cal{N}}$.
Since 
$\lambda_1(\mu=0)=\Phi$ 
is the golden number 
and $\lambda_2(\mu=0)=1-\Phi=-1/\Phi$
we may use the properties of powers of $\Phi$ and its reciprocal.
Namely, 
for any even integer $n$  $\Phi^n+1/\Phi^n$ is a whole number;
in particular, 
$2,\;3,\;7,\;18,\;47,\;123,\;322$ 
for 
$n=0,\;2,\;4,\;6,\;8,\;10,\;12$.
(For any odd integer $n$  $\Phi^n-1/\Phi^n$ is a whole number;
in particular, 
$1,\;4,\;11,\;29,\;76,\;199$ 
for 
$n=1,\;3,\;5,\;7,\;9,\;11$.)
See: http://goldennumber.net/phipower.htm.
\label{tab3}}
\vspace{5mm}
\begin{tabular}
{|c||c|c|c|c|c|c|} \hline
${\cal{N}}$ & ${\cal{W}}$ & sawtooth & two-leg ladder & two-leg ladder
                          & kagom\'{e}-like, & kagom\'{e}-like, \\ 
            &             &                & ($J_2=2J_1$)      & ($J_2=3J_1$) 
                          & Fig. \ref{fig1}e & Fig. \ref{fig1}f  \\ \hline \hline
 4  &   7 &   7 &   8 &   7 &   9 &   9  \\ \hline
 6  &  18 &  18 &  19 &  18 &  20 &  20  \\ \hline
 8  &  47 &  47 &  48 &  47 &  49 &   -  \\ \hline
10  & 123 & 123 & 124 & 123 &   - &   -  \\ \hline
12  & 322 & 322 & 323 & 322 &   - &   -  \\ \hline
\end{tabular}
\end{center}
\end{table}

Next, we discuss 
the results for  the magnetization $(Ns-\langle S^z\rangle)/{\cal{N}}$,
the entropy $S/k{\cal{N}}$ and 
the specific heat $C/k{\cal{N}}$ in dependence on $(h-h_1)/kT$
for spin systems belonging 
to the dimer universality class, 
i.e. for the sawtooth chain, the frustrated two-leg ladder  and the 
kagom\'{e}-like chains of type I and II, see
Figs.~\ref{fig5} and \ref{fig6}. 
Again, the universal formulas (\ref{11}) -- (\ref{15}) 
depend only on $(h-h_1)/kT$, i.e. they are identical for all temperatures 
$kT=0.01,\;0.1,\;0.2,\;0.3$ considered in Figs.~\ref{fig5} and \ref{fig6}. 
However, as mentioned already above, they are size dependent. 
Hence, we compare  the spin data for systems of finite size $N$ 
with the hard-core data of corresponding size $\cal N$ (cf. Table~\ref{tab1}) 
obtained from Eq.~(\ref{11}).  
Note, however, 
that the systems considered  in Fig.~\ref{fig5} and in Fig.~\ref{fig6}, left panel, 
correspond to ${\cal N}=10$, where the curves for the finite 
hard-dimer system are already close to those for the 
thermodynamic limit.   

The results shown in Fig.~\ref{fig5} belong to systems
with no extra spin states, 
i.e. ${\cal{W}}_{\rm{sp}}={\cal{W}}$, see Table~\ref{tab3}, 
and quite large energy separation $\Delta \approx 0.8$ 
(sawtooth chain, $N=18$) and   $\Delta \approx 0.9$ 
(two-leg ladder, $J_1=1$, $J_2=3$, $N=20$).
Consequently, we find an excellent agreement between the spin 
data and the hard-dimer data for temperatures up to $kT\approx 0.1 \approx 0.1 \Delta$.  
For larger temperatures again the qualitative behavior 
is well described by the universal formulas derived  from 
Eq.~(\ref{11}). 
The modified hard-core description based on 
Eq.~(18)
with appropriate parameters $U$ leads to a further  quantitative improvement, 
see Fig.~\ref{fig5}.   

Fig.~\ref{fig6} corresponds to spin systems, for which
the number of degenerate spin states  ${\cal{W}}_{\rm{sp}}$
at the saturation field for finite spin systems 
is larger
than the number of states ${\cal{W}}$ of corresponding finite hard-core object lattice gas,
see Table \ref{tab3}. 
However, we have only one (two-leg ladder, $J_1=1$, $J_2=2$) or two (kagom\'{e}-like chains I and II)  
extra spin states, which 
become irrelevant for larger $N$. 
The energy separation $\Delta$ 
for these systems with extra spin states we estimate to
$\Delta \approx 0.9$ 
for the ladder ($J_1=1$, $J_2=2$, $N=20$),
 $\Delta \approx 0.4$ for
the kagom\'{e}-like chain I ($N=18$), and  $\Delta \approx 0.2$ for
the kagom\'{e}-like chain II ($N=20$).
Therefore
we obtain (see Fig.~\ref{fig6})
a perfect agreement between the spin and the hard-core data only for very low temperatures, but 
realize a quantitative deviation  already for $kT\approx 0.1$.
Note that 
for the 
the kagom\'{e}-like chain II, the size $N=20$ of the spin system 
corresponds to only ${\cal{N}}=4$, and the finite size 
effects are clearly largest in this case.

\begin{figure}
\begin{center}
\includegraphics[clip=on,width=75mm,angle=0]{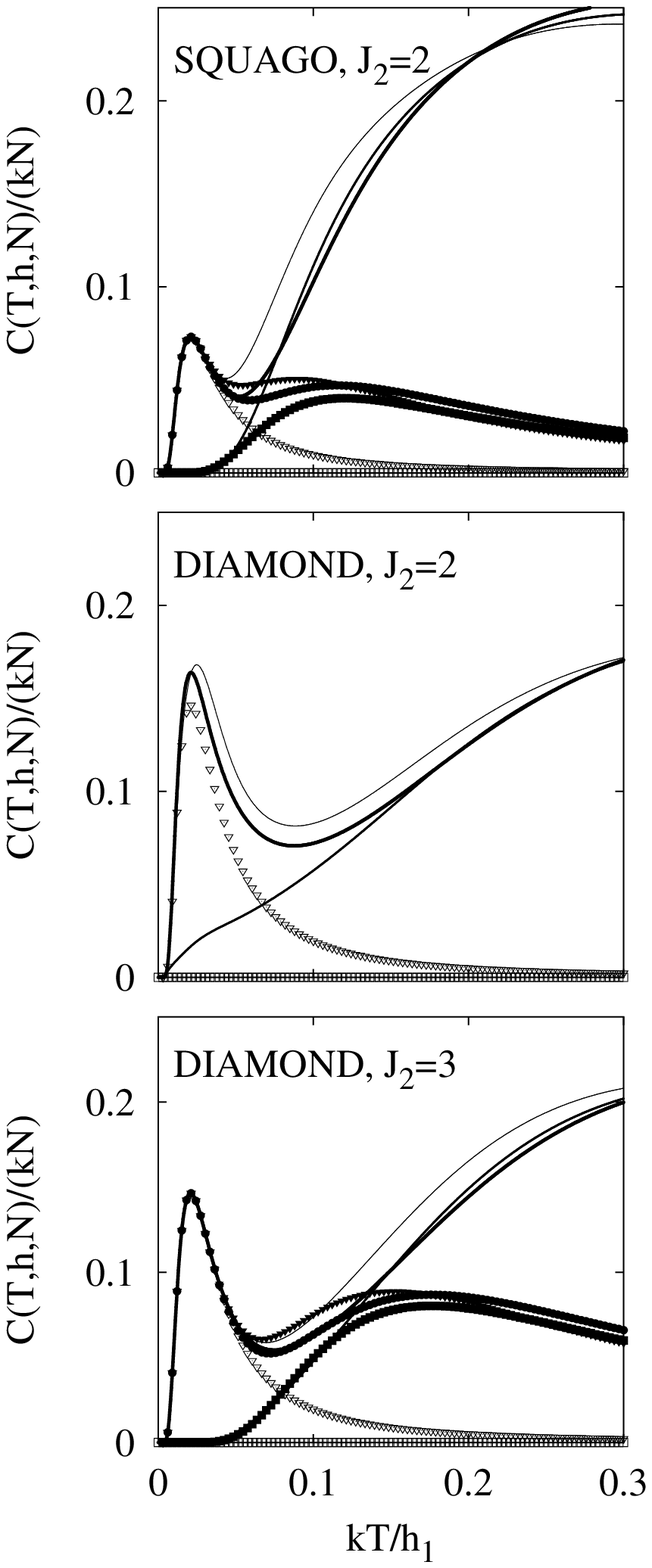}
\hspace{0mm}
\includegraphics[clip=on,width=75mm,angle=0]{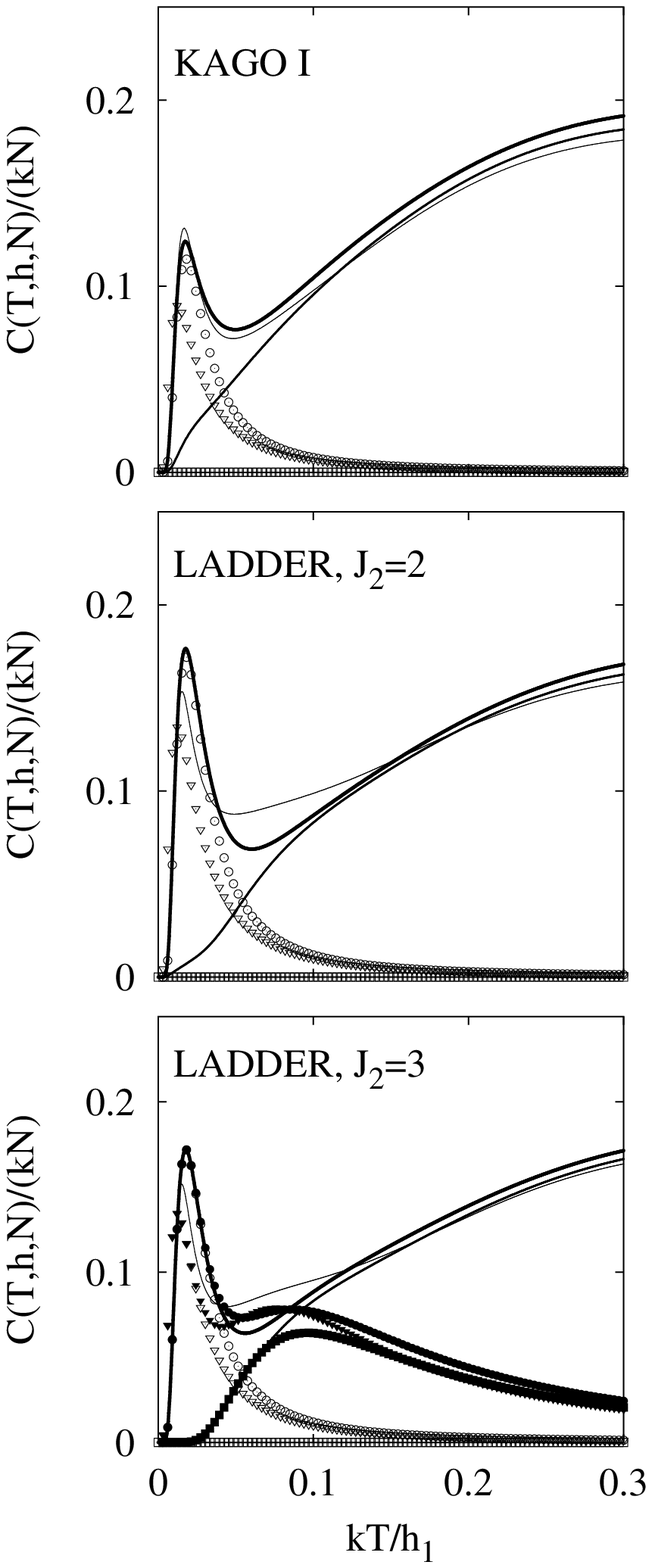}
\end{center}
\caption[]
{The temperature dependences of the specific heat 
$C/kN$ vs. $kT/h_1$ 
for several finite spin systems.
Monomer universality class:
left panels, from bottom to top,
the diamond chain with $J_1=1$, $J_2=3$, $N=18$,
the diamond chain with $J_1=1$, $J_2=2$, $N=18$,
the square-kagom\'{e} lattice $J_1=1$, $J_2=2$, $N=18$.
One-dimensional dimer universality class:
right panels, from bottom to top,
the two-leg ladder with $J_1=1$, $J_2=3$, $N=20$,
the two-leg ladder with $J_1=1$, $J_2=2$, $N=20$,
the kagom\'{e}-like chain shown in Fig. \ref{fig1}e with $J=1$, $N=18$.
The thin solid lines correspond to $h=0.95 h_1$,
the solid lines correspond to $h=h_1$,
the thick solid lines correspond to $h=1.05 h$.
For comparison we present analytical results 
for monomers (left panels) and dimers, ${\cal{N}}\to\infty$ (right panels) 
(empty symbols; 
triangles correspond to $h=0.95 h_1$,
squares correspond to $h=h_1$,
circles correspond to $h=1.05 h_1$).
We also show by filled symbols 
the temperature dependence of the specific heat 
as it follows 
from Eq. (\ref{10}) 
with $U=1.57$ for the diamond chain 
and $U=1.34$ for the square-kagom\'{e} lattice 
or 
from Eq. (\ref{18}) for ${\cal{N}}\to\infty$ 
with $U=1.10$ for the two-leg ladder.
\label{fig7}}
\end{figure}

\begin{figure}
\begin{center}
\includegraphics[clip=on,width=75mm,angle=0]{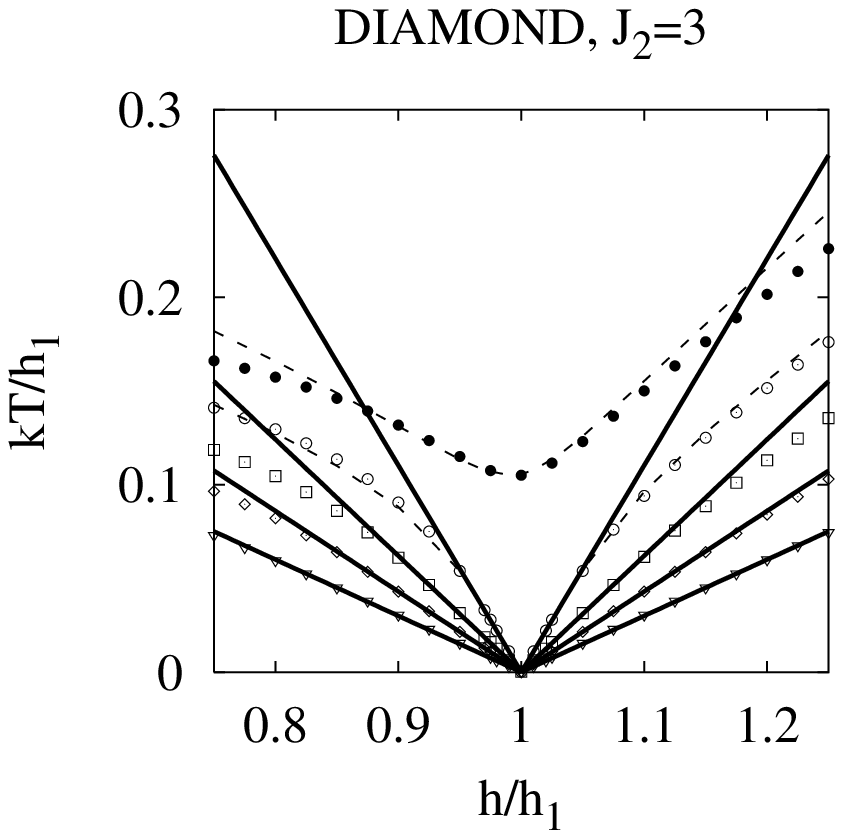}
\hspace{0mm}
\includegraphics[clip=on,width=75mm,angle=0]{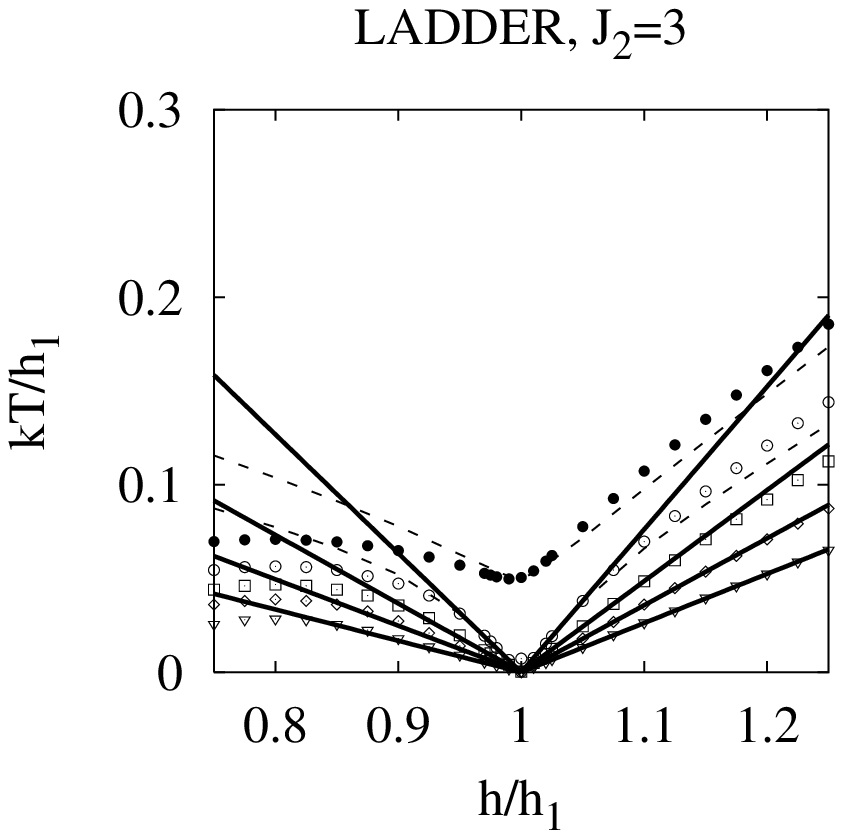}
\end{center}
\caption[]
{Constant entropy curves as a function of magnetic field and temperature, 
$S(T,h,N)/kN={\rm{const}}$,
for the diamond chain with $N=18$, $J_1=1$, $J_2=3$ (left)
and the two-leg ladder with $N=20$, $J_1=1$, $J_2=3$ (right).
From bottom to top: 
$S(T,h,N)/kN=0.05,\;0.10,\;0.15,\;0.20,\;0.25$.
Solid lines correspond to 
analytical results which follow from Eq. (\ref{07}) (left panel) 
or from Eq. (\ref{13}) (right panel);
broken lines correspond to 
analytical results for $S(T,h,N)/kN=0.20,\;0.25$
which follow from Eq. (\ref{10}) with $U=1.57$ (left panel) 
or from Eq. (\ref{18}) for ${\cal{N}}\to\infty$  with $U=1.10$ (right panel);
symbols correspond to exact diagonalization data for finite spin systems.
\label{fig8}}
\end{figure}

From the experimental point of view
the reported results manifest themselves most interestingly in a drastic 
change of the low-temperature specific heat, when the 
magnetic field passes the saturation field, and in the maximum 
of the isothermal entropy at saturation field leading to an enhanced 
magnetocaloric effect (for a general discussion of the magnetocaloric effect
for quantum spin systems, see Ref.~\cite{12}).  
We illustrate that in Figs.~\ref{fig7} and \ref{fig8}, where we present 
finite-lattice data for spin systems, but hard-core results 
for the thermodynamic limit to demonstrate that the 
discussed  effects remain relevant for $N \to \infty$.

In Fig.~\ref{fig7} we present 
the temperature dependence of the specific heat
at three values of the external magnetic field, 
$h=0.95 h_1, h_1, 1.05 h_1$, 
for several spin systems. 
Firstly, 
we note that within the hard-core object description 
the specific heat equals to zero at saturation field $h=h_1$. 
That is also observed  
in a certain range of low temperatures around zero
for the diamond chain with $J_2>2J_1$, 
the square-kagom\'{e} lattice with $J_2>J_1$
or 
the two-leg ladder with $J_2>2J_1$,
but we do not observe the zero-value region of $C/kN$ at low temperatures 
for the spin systems having extra spin states and low separation $\Delta$.
Secondly, 
we see that the specific heat 
 in the vicinity of the saturation field (but $h\ne h_1$) 
has an extra low-temperature peak 
which is satisfactorily reproduced within the hard-core 
object lattice gas approach
(compare lines and empty triangles and circles in Fig. \ref{fig7}).
The hard-core description can be improved assuming a finite on-site 
energy parameter $U$  
(filled triangles and circles in Fig. \ref{fig7}).
As it was already noticed in Ref. \cite{08} 
in the context of the sawtooth chain and the kagom\'{e} lattice, 
the experimental observation of the low-temperature peak of the specific heat 
can be a sign of highly degenerate localized magnon states.

Finally, we consider in Fig. \ref{fig8} adiabatic cooling processes, i.e. we
show curves    of 
 constant entropy as a function of magnetic field and temperature.
Again the exact diagonalization data 
for the diamond chain with $N=18$, $J_1=1$, $J_2=3$ (left panel)
and
the two-leg ladder with $N=20$, $J_1=1$, $J_2=3$ (right panel)
are in a good agreement 
with the universal curves for monomers 
and one-dimensional dimers.
Taking the data for spin systems, which show some asymmetry 
between $h>h_1$ and $h<h_1$, we observe, that an efficient
 cooling to very low temperatures can be achieved by an adiabatic 
demagnetization via lowering the magnetic field from above 
saturation till saturation.

\section{Summary}
\label{5}

We have presented  a universal description of thermodynamic properties 
of a wide class of frustrated quantum spin antiferromagnets   
at low temperatures in the vicinity of the saturation field.
The reason for that is a dominant contribution of highly degenerate localized magnon states 
to the partition function, which can be calculated 
using a hard-core object representation. 
We have provided exact diagonalization data 
to justify such a picture. 

The spin systems hosting localized magnons can be grouped into different universality 
classes of hard-core lattice gases, 
namely gases of hard squares, hard hexagons, hard dimers or hard monomers.
Comparing the analytical predictions 
with numerical data for finite spin systems 
we have found 
that 
the hard-core picture describes accurately  
the spin physics 
near the saturation field and at sufficiently low temperatures. 
Since 
other states of the spin system, 
not described by the hard-core model, 
become relevant as temperature grows, 
the hard-core description is less accurate, 
but it remains qualitatively correct at higher temperatures. 
The hard-core object lattice gas models may be improved 
by  relaxing the hard-core constraint.
Such an improvement breaks the  universality, 
but provides a better quantitative agreement with exact diagonalization results 
for higher temperatures
and larger deviations from the saturation field. 

We emphasize 
that some peculiarities of the thermodynamic quantities 
arising due to the localized magnon states
are of interest from the experimental point of view.
Thus,
the specific heat at the saturation field 
remains almost zero below a certain temperature
whereas at fields  
slightly above/below the saturation field
the specific heat exhibits a well-pronounced extra low-temperature peak.
A frustrated spin system hosting localized magnons also exhibits 
a large magnetocaloric effect 
in the vicinity of the saturation field;
similarly to an ideal paramagnet.

\section*{Acknowledgments}

We thank T.~Krokhmalskii 
for discussions on two-dimensional lattice gases of hard-core objects 
and A.~Honecker and J.~Schnack for a critical reading of the manuscript. 
The numerical calculations were performed using J.~Schulenburg's {\it spinpack}.
We also thank the DFG for the support 
(Project No. 436 UKR 17/13/05).
One of the authors (O.~D.) thanks 
the Abdus Salam International Centre for Theoretical Physics at Trieste, 
where a part of this work was carried out, 
for hospitality in the autumn of 2005.

\end{document}